\documentclass{article}
\usepackage{emulateapj}
\usepackage{apjfonts}

\begin{document}

\submitted{To appear in The Astronomical Journal}

\title{RR Lyrae Stars in the Andromeda Halo from Deep Imaging with the 
Advanced Camera for Surveys$^1$}

\author{Thomas M. Brown, Henry C. Ferguson, Ed Smith}

\affil{Space Telescope Science Institute, 3700 San Martin Drive,
Baltimore, MD 21218;  tbrown@stsci.edu, ferguson@stsci.edu,
edsmith@stsci.edu} 

\medskip

\author{Randy A. Kimble, Allen V. Sweigart}

\affil{Code 681, NASA Goddard Space Flight Center, Greenbelt, MD 20771; 
randy.a.kimble@nasa.gov, allen.v.sweigart@nasa.gov} 

\author{Alvio Renzini}

\affil{European Southern Observatory, Karl-Schwarzschild-Strasse 2, 
Garching bei M$\ddot{\rm u}$nchen, Germany; arenzini@eso.org} 

\author{R. Michael Rich}

\affil{Division of Astronomy, Dpt.\ of Physics \& Astronomy, UCLA, Los
Angeles, CA 90095; rmr@astro.ucla.edu}

\begin{abstract}

We present a complete census of RR Lyrae stars in a halo field of the
Andromeda galaxy.  These deep observations, taken as part of a program
to measure the star formation history in the halo, spanned a period of
41 days with sampling on a variety of time scales, enabling the
identification of short and long period variables.  Although the long
period variables cannot be fully characterized within the time span of
this program, the enormous advance in sensitivity provided by the
Advanced Camera for Surveys on the Hubble Space Telescope allows
accurate characterization of the RR Lyrae population in this field.
We find 29 RRab stars with a mean period of 0.594 days, 25 RRc stars
with a mean period of 0.316 days, and 1 RRd star with a fundamental
period of 0.473 days and a first overtone period of 0.353 days.  These
55 RR Lyrae stars imply a specific frequency $S_{RR}\approx 5.6$,
which is large given the high mean metallicity of the halo, but
not surprising given that these stars arise from the old, metal-poor
tail of the distribution. This old population in the Andromeda halo
cannot be clearly placed into one of the Oosterhoff types: the ratio
of RRc/RRabc stars is within the range seen in Oosterhoff II globular
clusters, the mean RRab period is in the gap between Oosterhoff types,
and the mean RRc period is in the range seen in Oosterhoff I globular
clusters.  The periods of these RR Lyraes suggest a mean metallicity
of [Fe/H]$\approx-1.6$, while their brightness implies a distance
modulus to Andromeda of 24.5$\pm 0.1$, in good agreement with the
Cepheid distance.

\end{abstract}

\keywords{galaxies: evolution -- galaxies: stellar content --
galaxies: halos -- galaxies: individual (M31) -- stars: variables}

\section{Introduction}

The textbook picture of a spiral galaxy halo comes from that of our
own Milky Way, which is old and metal-poor (VandenBerg 2000; Ryan \&
Norris 1991).  However, the stellar population of the Andromeda (M31;
NGC224) halo offers a striking contrast to this picture, with its wide
range in metallicity (Durrell, Harris, \& Pritchet 2001) and age
(Brown et al.\ 2003; Brown 2003).  These spreads in metallicity and age can
significantly affect the variable star population.  For example, the
characteristics of RR Lyraes in Galactic globular clusters place these
clusters into two distinct Oosterhoff types (Oosterhoff 1939), while
the RR Lyraes in Local Group dwarf spheroidals (dSphs) place these
galaxies in the gap between the two Oosterhoff types (Siegel \& Majewski 2000;
Dall'Ora et al.\
2003; Pritzl et al.\ 2002, 2004).  Compared to the M31 halo, dSphs have an
even broader age range (van den Bergh 1999) and are generally more
metal-poor (Mateo 1998), but the old ($> 10$~Gyr) component capable of
producing RR Lyrae stars might be similar in each case.

The specific frequency of RR Lyraes in the M31 halo has been the
subject of some debate.  In a field 40 arcmin from the \linebreak

{\small \noindent $^1$Based on observations made with the NASA/ESA
Hubble Space Telescope, obtained at the Space Telescope Science Institute, 
which is operated by AURA, Inc., under NASA contract NAS 5-26555. These
observations are associated with proposal 9453.}

\noindent
nucleus on the
southeast minor axis, Pritchet \& van den Bergh (1987) found 30 RR
Lyraes, and with an estimated completeness of 25\%,
determined that the frequency per unit luminosity was very high (about
half of that in variable-rich M3).  More recently, Dolphin et al.\
(2003) found only 24 RR Lyraes in a larger field that included the
Pritchet \& van den Bergh (1987) field, and with their estimated
completeness of 24\%, claimed that the frequency of RR Lyrae
was $\sim$15 times smaller.  To support their claim, they estimated
that the deep color magnitude diagram (CMD) of Brown et al.\ (2003)
contained only 10 RR Lyraes, but as we shall show here, this was a
severe underestimate.

We have observed a field along the southeast minor axis of the M31
halo, 51 arcmin from the nucleus, using the Advanced Camera for
Surveys (ACS; Ford et al. 1998) on the Hubble Space Telescope (HST).
The primary goal of this program was to investigate the halo star
formation history, by constructing a deep CMD
in the F606W (broad $V$) and F814W ($I$) bandpasses, reaching
$V\approx 30.7$~mag on the main sequence (Brown et al.\ 2003).
However, the 250 individual exposures are scattered with variable time
sampling over a 41 day period, and thus provide excellent time series
photometry for the variable star population in the M31 halo; in
particular, the completeness for RR Lyraes in our field is
approximately 100\%.  In this paper, we present a survey of the RR Lyraes
and the other bright variables in our field.

\section{Observations and Data Reduction}

Using the Wide Field Camera on the ACS, we obtained deep optical
images of a $3.5\arcmin \times 3.7\arcmin$
field along the southeast minor axis of the M31 halo, at
$\alpha_{2000} = 00^h46^m07^s$, $\delta_{2000} = 40^{\rm
o}42^{\prime}34^{\prime\prime}$.  The surface brightness in this region
is $\mu_V \approx 26.3$~mag arcsec$^{-2}$ (Brown et al.\ 2003).  The
area is not associated with the tidal streams and substructure found
by Ferguson et al.\ (2002), and lies just outside the ``flattened
inner halo'' in their maps.  We placed a metal-rich M31 globular cluster, 
GC312 (Sargent et al.\ 1977), near the edge of the field.  From 2
Dec 2002 to 11 Jan 2003, we obtained 39.1 hours of images in the F606W
filter (broad $V$) and 45.4 hours in the F814W filter ($I$), with each
of the 250 exposures dithered to allow for hot pixel removal, optimal
point spread function sampling, smoothing of spatial variations in
detector response, and filling in the gap between the two halves of
the $4096 \times 4096$ pixel detector.  Of the 250 exposures, 16 were
short ($< 600$~s), to allow the correction of bright saturated
objects, leaving 234 long exposures ($> 1200$~s) suitable for deep 
time-series photometry.

In order to create a deep catalog of the objects in this field, we
first co-added these images using the IRAF DRIZZLE package.  This
coaddition included masks for the cosmic rays and hot pixels, and
produced geometrically-correct images with a plate scale of
0.03$^{\prime\prime}$ pixel$^{-1}$ and an area of approximately
$210^{\prime\prime} \times 220^{\prime\prime}$.  The mask for each
exposure was created in an iterative process, comparing the value in
every pixel to the distribution through the entire stack at that
location on the sky; pixels are masked on the bright end of the
distribution (cosmic rays and hot pixels) and the faint end of the
distribution (dead pixels), but this technique also masks bright,
large-amplitude, variable stars when they are near maximum or minimum.
The final coadded image in each bandpass thus approximates the average
flux observed for the variable stars, but these masks cannot be used
in the time series photometry (discussed below).

We then performed both aperture and PSF-fitting photometry using the
DAOPHOT-II package (Stetson 1987), assuming a variable PSF constructed
from the most isolated stars.  The aperture photometry on isolated
stars was corrected to true apparent magnitudes using TinyTim models
of the HST PSF (Krist 1995) and observations of the standard star EGGR
102 (a $V=12.8$~mag DA white dwarf) in the same filters, with
agreement at the 1\% level.  The PSF-fitting photometry was then
compared to the corrected aperture photometry, in order to derive the
offset between the PSF-fitting photometry and true apparent
magnitudes. Our photometry is in the STMAG system: $m= -2.5 \times
$~log$_{10} f_\lambda -21.1$.  For readers more familiar with the
Johnson $V$ and Cousins $I$ bandpasses, a star in the middle of the RR
Lyrae strip has $V - m_{F606W} = -0.17$~mag and $I - m_{F814W} =
-1.29$~mag.  Light curve amplitudes are typically 6--10\% smaller
in $m_{F606W}$ than in $V$, and 1--2\% smaller in $m_{F814W}$ than in $I$.
The transformation between the ACS and ground-based
bandpasses is still being characterized independently by several
groups, so for most of this paper we will refer to magnitudes in
the unambiguous STMAG system.  

The CMD for the co-added images was shown by Brown et al.\ (2003).  In
that analysis, we discarded $\approx$20\% of the exposed area (around
bright foreground stars, near GC312, and in regions with
less than the full exposure time due to the dither pattern).  In
the current analysis of the brighter stars, we include the entire
image area.  We will only consider stars brighter than $m_{F814W} =
28.25$~mag, which have a signal-to-noise ratio of $\sim 5$ in
individual exposures.  Extensive artificial star tests demonstrate
that our catalog, created from the deep co-added data, is
$\approx$100\% complete above this limit. 

To obtain time-series photometry, we re-drizzled the entire dataset
into a stack of individual registered exposures, with masked pixels
set to an invalid data value.  Because our original masks sometimes
discard data points near the maxima and minima of variable stars, we
unmasked those pixels that were masked more than once in a two-hour
window; this correction will occasionally restore a true cosmic ray or
dead pixel when it should have been masked, but these occasional
events will be significant outliers in the time series photometry,
which can be discarded after the fact.  We then performed aperture
photometry on each individual frame, with positions fixed by the
catalog of the coadded exposures.

\section{Variable Detection and Characterization}

As explained above, we restricted our time-series analysis to stars
brighter than $m_{F814W} = 28.25$~mag; out of the nearly 300,000 stars
in the full image catalog, these 19,450 stars have a signal-to-noise
of at least 5 in individual exposures.  Stars in the RR Lyrae gap,
near $m_{F814W} \approx 26$~mag, have typical photometric errors of
0.03~mag in F606W exposures and 0.04~mag in F814W exposures.  Most of
the stars in our search list have $\sim 100$ photometric measurements
in each bandpass over our 41 day observing program, but a small
fraction have significantly less because they fell near the edges of
the two halves of the detector; 424 stars, representing only 2\% of the
stars in the relevant brightness range, were discarded because they
had less than 30 valid measurements in each bandpass. For the remaining
stars, we looked for variability using two methods, chosen to suit the
high signal-to-noise and large number of time samples in these data.

The first variability search was tuned to provide a complete sample of
RR Lyraes, but it also recovered a large number of brighter, long
period variables and fainter, short period variables.  We used a fast
algorithm (Press \& Rybicki 1989) of the Lomb-Scargle periodogram
(Lomb 1976; Scargle 1982), which looks for weak periodic signals in
irregularly sampled data.  Besides providing a good initial estimate
of the period, this algorithm also quantifies the statistical
significance of the periodic signal.  Our threshold was a non-random
signal at 0.01 significance, independently found in each bandpass; a
score of 0.01 implies that the chance this signal arose from random
fluctuations is less than 1 percent.  This stringent threshold should
only produce one or two false detections in the entire search list.
However, this threshold is not so stringent that it misses RR Lyraes;
all of the RR Lyraes recovered by this method, regardless of amplitude
or period, were detected at a significance orders of magnitude beyond
this threshold (if our choice of threshold were in fact discarding RR
Lyraes, then the distribution of RR Lyrae scores would approach the
threshold).  The initial period estimate returned by this method was
then refined by a search of the Lafler \& Kinman (1965) statistic, for
periods within 5\% of the initial period.  Our method recovered 169
variables, two of which were false detections (each a faint star
sitting in the wings of a bright variable), and 55 of which are
clearly RR Lyraes.  Of the remaining variables, 17 are short period
variables below the horizontal branch (HB), 3 have periods of 0.6--7
days and lie above the HB, 1 is an eclipsing binary, 82 are long
period variables (LPVs) and semiregulars near the tip of the red giant
branch (RGB), 5 are LPVs fainter than the HB, and 4 are variables with
periods of $\sim$1 week that lie on the RGB.  Thirteen of the detected
variables, all with periods \linebreak

\hskip 0.2in
\epsfxsize=6.5in \epsfbox{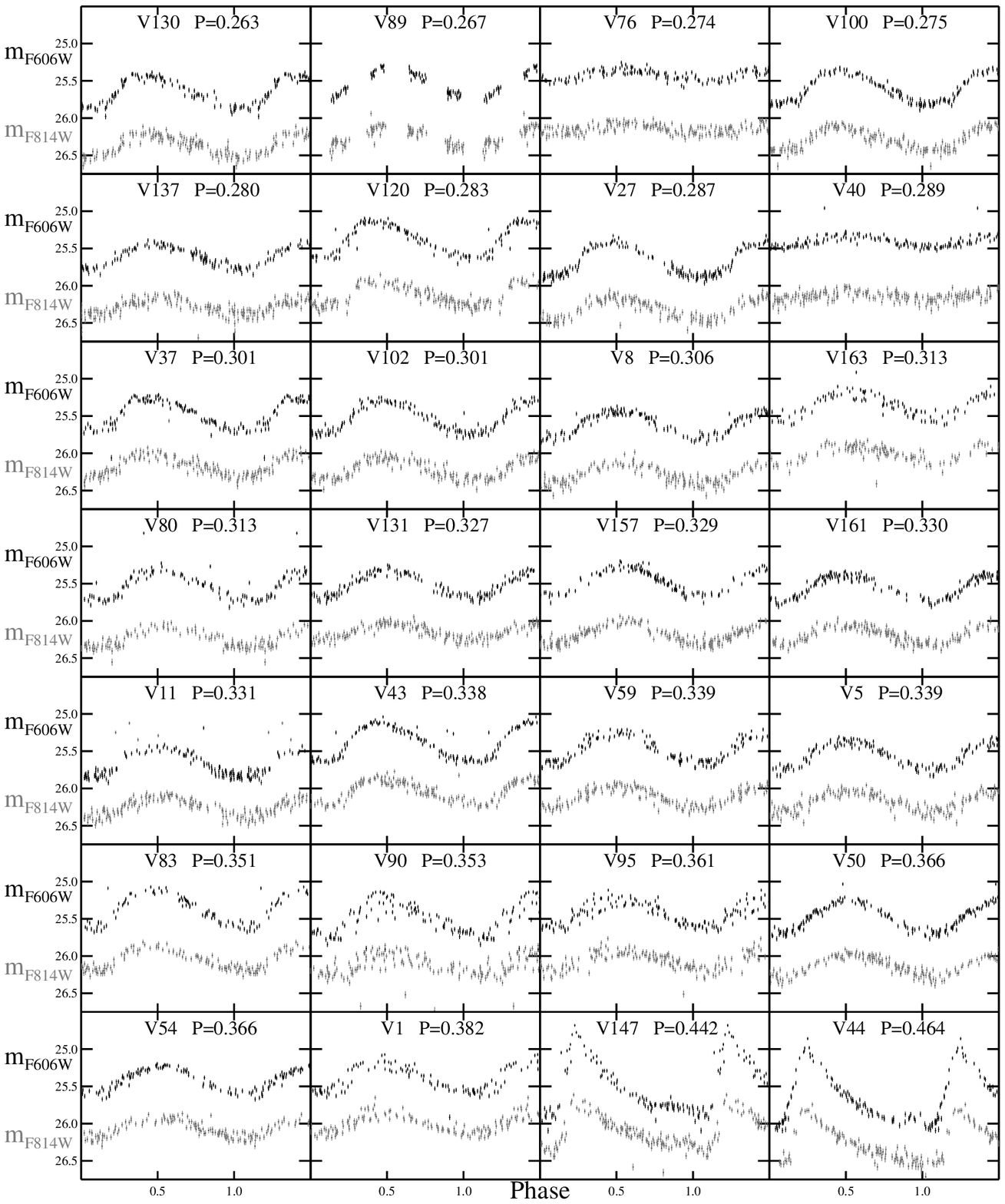}

\hskip 0.2in
\parbox{6.5in}{\small {\sc Fig.~1--}
The light curves for RR Lyraes in our field, arranged
in order of increasing period ({\it labeled}), shown as photometric
error bars (not data points).  The first 26 stars are
RRc and RRd stars (V90 being the RRd, shown phased with its first
overtone period); the remaining 29 are RRab stars.}

\vskip 0.1in

\noindent
longer than 20 days, fall within the tidal
radius of GC312 (10$\arcsec$; Holland et al.\ 1997); because the area
within the tidal radius of GC312 comprises less than 1 percent of our
total field, these 13 LPVs and semiregulars are clearly associated
with GC312.

The second variability search simply looked for photometry that
exceeded the expected scatter (given the photometric \linebreak

\vskip 8.5in

\noindent
errors) by 50\%.
This method was motivated by a search for halo-on-halo microlensing
events.  We found no microlensing events or new pulsating variables,
but we did find 7 additional eclipsing binaries.  As might be expected,
the Lomb-Scargle periodogram is better suited to finding variables
with roughly sinusoidal light curves.  We show the light curves for
the RR \linebreak

\hskip 0.2in
\epsfxsize=6.5in \epsfbox{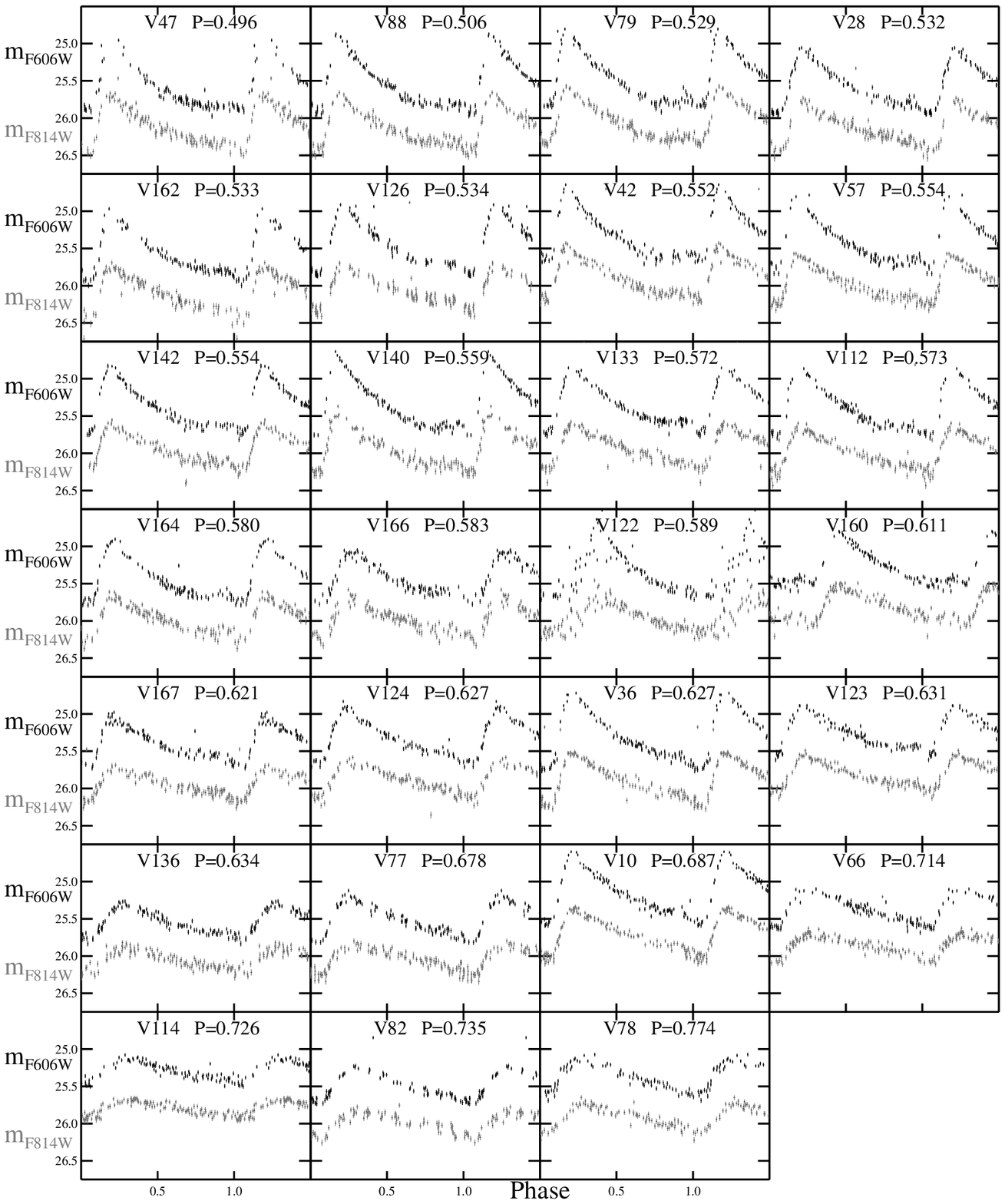}

\hskip 0.2in
\parbox{6.5in}{\small {\sc Fig.~1--} Continued.}

\vskip 0.1in

\noindent
Lyraes in Figure 1, light curves for the variables above and
below the HB in Figure 2, light curves for the eclipsing binaries in
Figure 3, and light curves for a subset of the LPVs and semiregulars
in Figure 4 (only those that varied by more than $\sim 0.1$~mag within
the 41 days of observations).  Note that only photometric error bars
are visible at the plotted scale of these light curves.

Five of the RR Lyrae light curves show significant scatter \linebreak

\vskip 8.28in

\noindent
(V1, V90, V95,
V122, and V163).  We reevaluated the periodicity searches on these five
stars, and found that one of the stars, V90, showed clear evidence for
double-mode pulsations, making it a definite RRd star, with a
fundamental period ($P_0$) of 0.4735 days and a first overtone period
($P_1$) of 0.3534 days.  V163 shows a hint of an additional period
at 0.4243 days, but the signal is weak, and we found no additional
periods \linebreak

\hskip 0.2in
\epsfxsize=6.5in \epsfbox{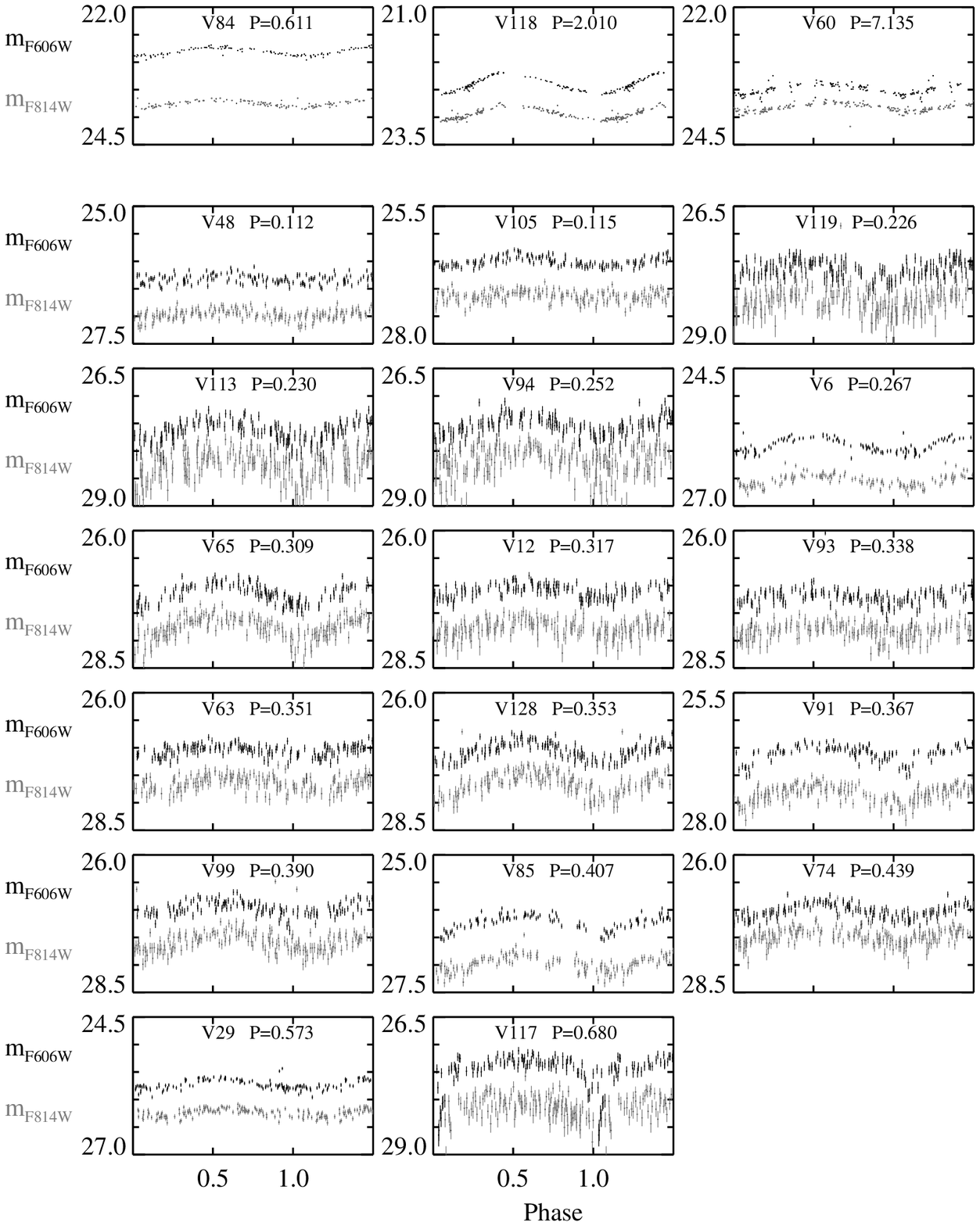}

\hskip 0.2in
\parbox{6.5in}{\small {\sc Fig.~2--}
The light curves for other variables bluer than the RGB
(excluding eclipsing binaries), shown as error bars.  The
stars in the top 3 panels are brighter than the HB; they are arranged
in order of increasing period and increasing $m_{F606W}-m_{F814W}$
color, making them easy to identify in the CMD (Figure 5).  The stars
in the lower 17 panels are likely dwarf Cepheids fainter than the HB,
but V6 and V29, which lie immediately below the HB, might be RR
Lyraes.}

\vskip 0.15in

\noindent
in the light curves of V1 and V95.  Although V1, V95, and V163
might be RRd stars, we
have classified them as RRc.  The scatter in the light curve of V122
might be due to the Blazhko effect -- a secondary modulation in the
variability of 20--30\% of RRab stars, with a period of $\sim$10--500
days (Smith et al.\ 2003). \\

\vskip 8.65in

To determine the mean magnitudes and amplitudes of the variables with
periods less than $\sim 10$ days, we fit a Fourier series to the
magnitudes in each bandpass as a function of phase, with an order of 1
or 2 for the sinusoidal light curves (e.g., RRcd stars) and an order
of 8 for the sawtooth light curves (e.g., \linebreak

\hskip 0.2in
\epsfxsize=6.5in \epsfbox{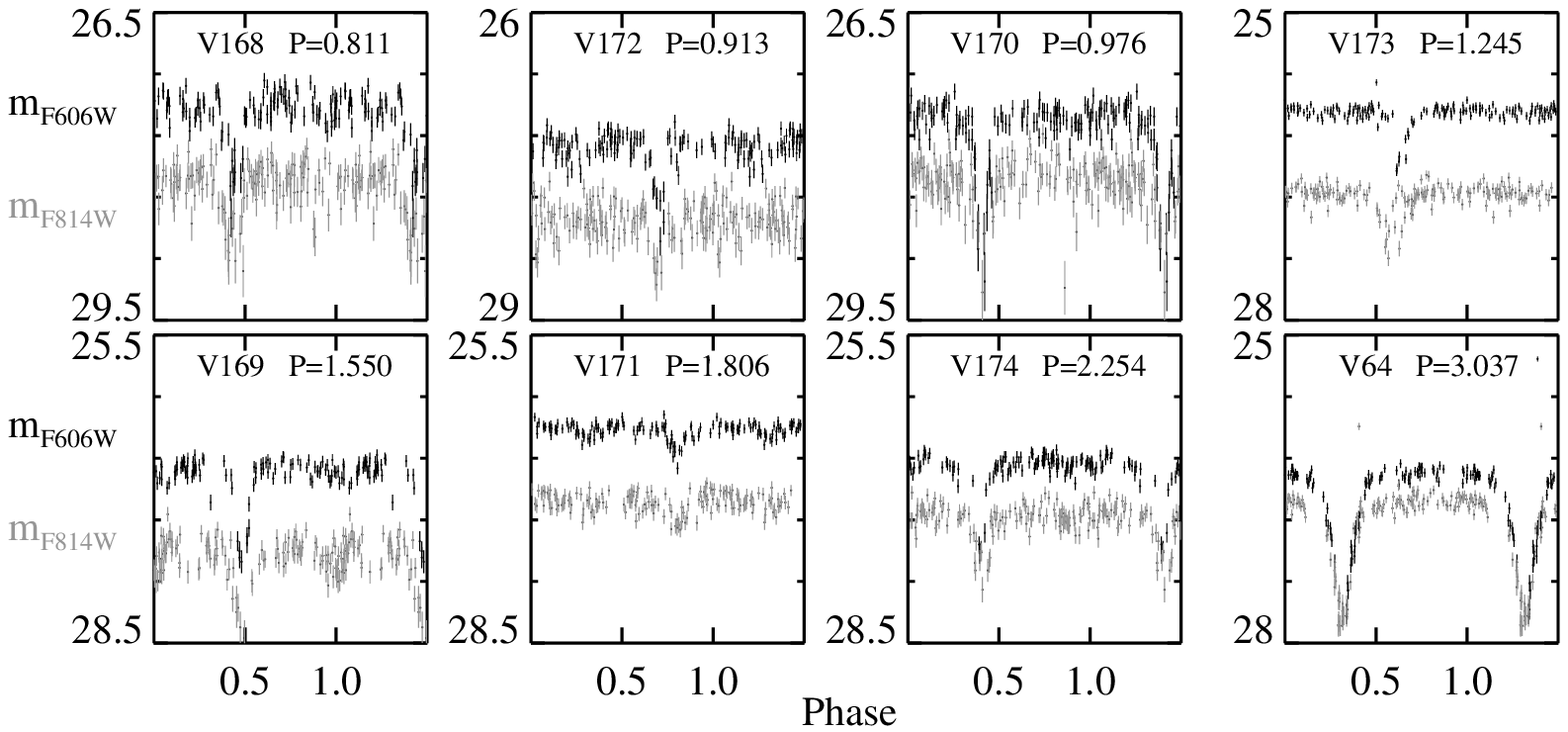}

\vskip 0.05in

\hskip 0.2in
\parbox{6.5in}{\small {\sc Fig.~3--}
The light curves for candidate eclipsing binaries in our
field.  The most secure (V64) was found through the Lomb-Scargle
periodogram search, while the others were found because their photometric
errors were larger than expected.  Note that a secondary minimum is
sometimes visible at a phase opposite to the primary minimum.}

\vskip 0.1in

\noindent
RRab stars).  Table 1 lists
the properties of the RR Lyraes in our field, and Table 2 gives their
time-series photometry (with the full dataset available only as
a machine-readable table in the electronic version of this paper).  
Table 3 gives the positions of the other variables in our field 
(Figures 2--4).  Figure 5
shows the CMD locations of the stars with light curves (Figures 1--4).
The variables are shown at their mean
$m_{F606W}-m_{F814W}$ color and $m_{F814W}$ magnitude, as determined
from the fitting above, instead of the average {\it observed} value in
the catalog from the deep co-added images, which can be systematically
brighter or fainter in a given bandpass by a few hundredths of a
magnitude, depending upon the random sampling of each light curve. 
Figure 6 shows the
distribution of $m_{F606W}$ amplitude vs. period, mean $m_{F606W}$ vs.
period, and the period distribution for the RR Lyrae stars.  The
RR Lyraes are cleanly separated from the remaining variables by their
position in the $m_{F606W}$ vs. period diagram, although two stars
lying immediately below the HB might also be RR Lyraes (V6 and V29
in Figure 2).  The RR Lyraes can be classified as either RRab or RRcd
by the gap in amplitude vs. period distribution; these classes do not
overlap in the CMD (Figure 5).

\section{RR Lyrae Properties}

\subsection{Oosterhoff Type}

The RR Lyrae population of the M31 halo cannot be classified into
either of the Oosterhoff (1939) types, but this is not because the various
characteristics appear as an average of the two types (Figure 7).
Looking at the ensemble variable population in Galactic globular
clusters, the ratio of RRc to RRabc stars is 0.22 in Oosterhoff I
clusters and 0.48 in Oosterhoff II clusters (Clement et al.\ 2001),
while in the M31 halo it is 0.46 -- a value in the Oosterhoff II
regime.  In contrast, looking at the mean period of the RRc stars in
Galactic globular clusters, it is 0.326 days in Oosterhoff I clusters
and 0.368 days in Oosterhoff II clusters (Clement et al.\ 2001), while
in the M31 halo it is 0.316 days, within the Oosterhoff I regime.  The
mean period for the known RRab stars in Oosterhoff I clusters is 0.559
days, and in Oosterhoff II clusters it is 0.659 days (Clement \linebreak

\vskip 3.5in

\noindent
et al.\
2001), while in the M31 halo it is 0.594 days -- midway between the
two types. The RRab period distribution is more sharply peaked at this
intermediate value than one might expect from an equal mix of
Oosterhoff types (see Figure 8), but a Kolmogorov-Smirnov test shows the
difference is not statistically significant.  The period-amplitude
diagram for the RRab stars 
is also suggestive of an intermediate Oosterhoff type (Figure 9).  
If we convert our $m_{F606W}$
amplitudes to $V$ amplitudes, and compare against the period-amplitude
relations of Oosterhoff I and Oosterhoff II clusters (Clement 2000),
the RRab stars in M31 fall between the two types, although there
is a tendency toward Oosterhoff I at the larger amplitudes.  

Although the Galactic halo field is more difficult to characterize
than the Galactic clusters, it is interesting to note that the field
RR Lyraes also show the Oosterhoff dichotomy, with a real gap in the
distribution of period shifts at fixed amplitude with respect to the
M3 RR Lyraes (see, e.g., Figure 8 of Suntzeff, Kinman, \& Kraft 1991).
However, these same data show that the Galactic field population is
predominantly of Oosterhoff type I, as has been known for some time.
In their study of the Palomar-Groningen Survey, Cacciari \& Renzini
(1976) found $<$$P_{ab}$$> = 0.529$~days, $<$$P_{c}$$> = 0.329$~days,
and $N_c / N_{abc} = 0.09$.  Selection effects in a field survey might
favor RR Lyraes with higher amplitudes (thus reducing the relative
number of RRc stars and long-period RRab stars), but these
characteristics are all well within the Oosterhoff I regime.  In
contrast, the corresponding values for the M31 halo RR Lyraes are
quite different.  In particular, the mean RRab period is much longer
in M31 ($<$$P_{ab}$$> = 0.594$~days), and the frequency of RRc
variables is much higher ($N_c / N_{abc} = 0.46$).  Most importantly,
the distribution of period shifts of the M31 RR Lyraes with respect to
the M3 RR Lyraes (i.e., the Oosterhoff I line in Figure 9) shows no
such gap.   All of these characteristics clearly distinguish the M31 
halo field from the Milky Way halo field.

The characteristics of the M31 halo variables do not track those of
the Local Group dSphs, either, except in the broad sense that the M31 halo
cannot be put cleanly into one of the Oosterhoff types (Figure 7).
For dSphs, the mean period of RRab stars tends to fall in the gap
between Oosterhoff types \linebreak

\hskip 0.2in
\epsfxsize=6.5in \epsfbox{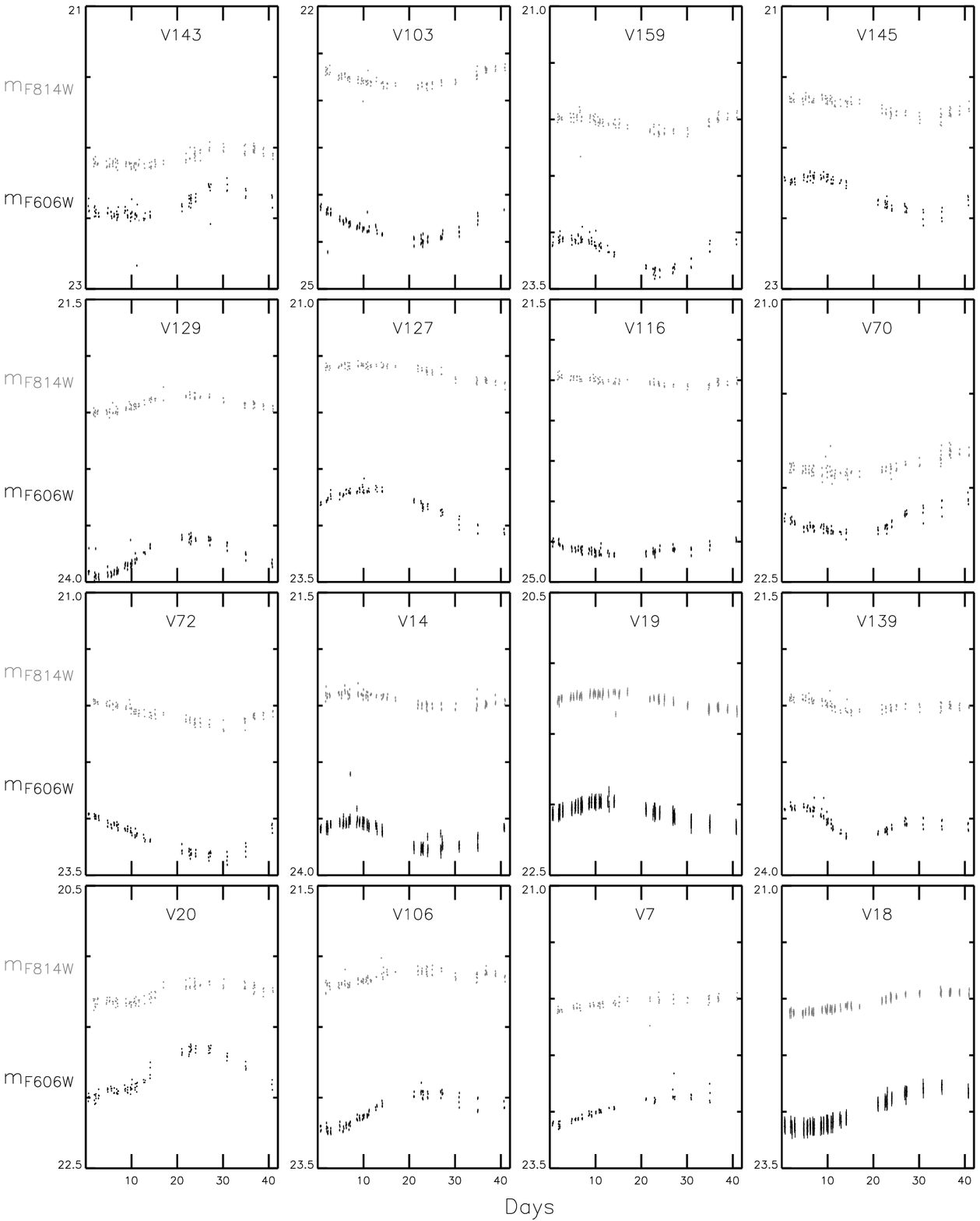}

\hskip 0.2in
\parbox{6.5in}{\small {\sc Fig.~4--}
The light curves for a subset of the bright LPVs and
semiregulars in our field
(those that varied by more than 0.1~mag in each bandpass), arranged
roughly in order of increasing period (the 41 day span of our observations
limits our ability to characterize the periods).  }

\vskip 0.1in

\noindent
(Siegel \& Majewski 2000; 
Pritzl et al.\ 2002, 2004; Dall'Ora et al.\ 2003
and references therein), as found in M31.  However, the fraction of RR
Lyrae stars that are RRc is higher in M31 (typical of Oosterhoff II)
than that in any of the dSphs (which tend toward the Oosterhoff I
values), while the mean period of RRc stars is much lower in M31
(typical of Oosterhoff I) than in any \linebreak

\vskip 8.47in

\noindent
of the dSphs (which tend to fall
in the gap between Oosterhoff types).  These differences might suggest
that the M31 halo is not comprised of dissolved
globular clusters like those in the Milky Way, or dissolved Local Group dSphs.
Note that globular clusters in the Large Magellanic Cloud are also 
predominantly of intermediate Oosterhoff type, in contrast to the 
clusters of the \linebreak

\hskip 0.2in
\epsfxsize=6.5in \epsfbox{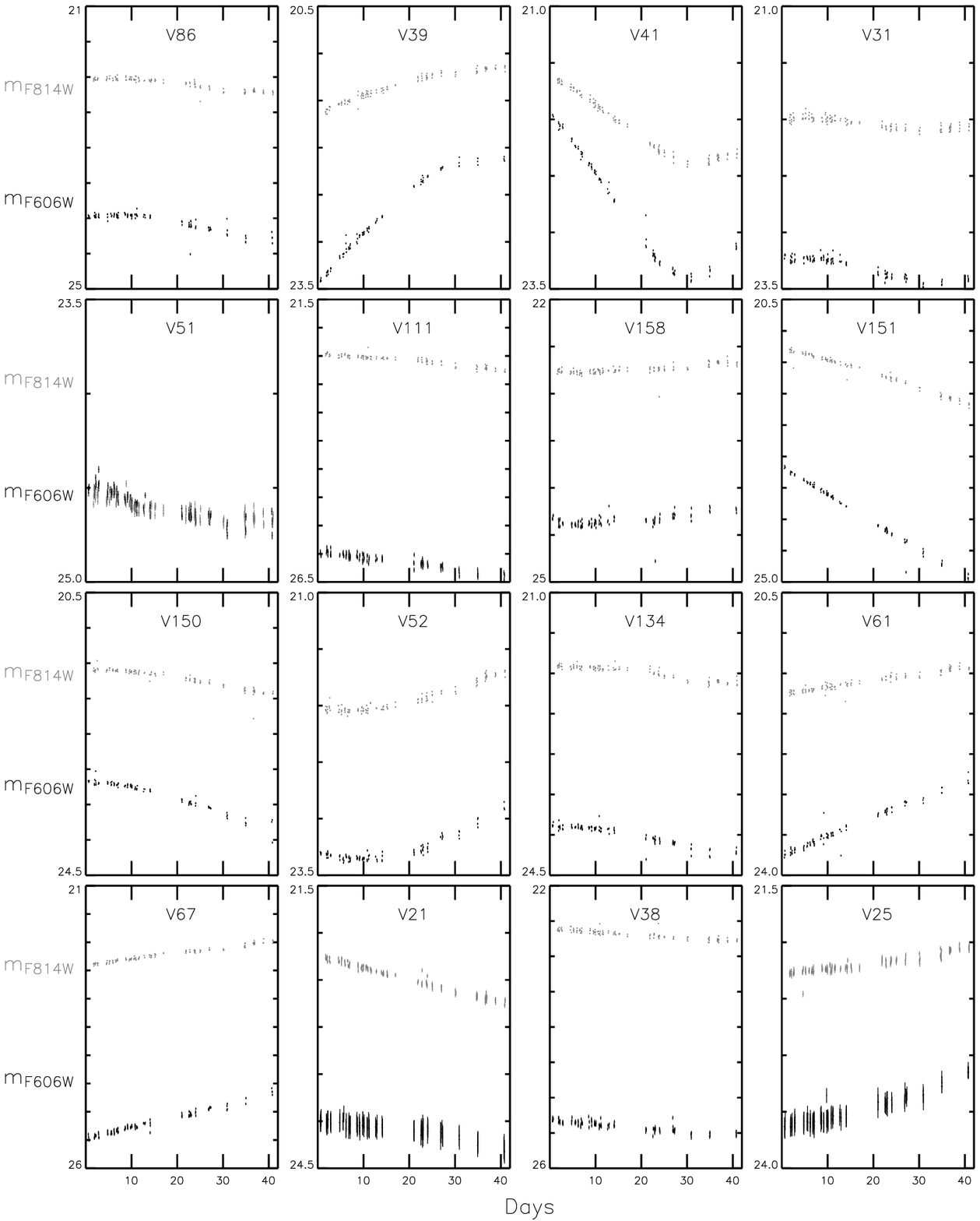}

\hskip 0.2in
\parbox{6.5in}{\small {\sc Fig.~4--} Continued.}

\vskip 0.09in

\noindent
Milky Way (Bono, Caputo, \& Stellingwerf 1994).

It has been argued that the Oosterhoff dichotomy among Galactic
globular clusters is due to a real gap in metallicity between the two
Oosterhoff types, with clusters at metallicities between the two Oosterhoff
types having very blue HBs that do not populate the RR Lyrae
instability strip (Renzini 1983; Sandage 1993).  Local Group dSphs
and LMC globular clusters fill the gap in a plot of
[Fe/H] vs.\ $<$$P_{ab}$$>$, as illustrated 
\linebreak

\vskip 8.26in

\noindent
by Siegel \& Majewski (2000) and Pritzl et al.\ (2004) 
(see Figure 6 in each paper).  This
indicates that the dSphs and Galactic globular clusters occupy
separate regions of the age vs. chemical composition parameter space (Renzini
1980), which in the simplest option reduces to the [Fe/H], age, and
[$\alpha$/Fe] three-dimensional space.    
The M31 halo appears to bridge the gap in a manner similar to
the dSphs, signaling that the early star formation history in the M31
halo proceeded at a different pace \linebreak

\hspace{-3.25in}
\parbox{6.0in}{
\begin{center}
{\sc Table 1 \\ RR Lyrae Properties}

\begin{tabular}{ccccccccc}
\tableline
     &     R.A.     & Dec.         & Period & $<$$m_{F606W}$$>$ & $<$$m_{F814W}$$>$ & 
$A_{F606W}$ & $A_{F814W}$ & \\
Name &      (J2000) &      (J2000) & (days) & (mag)         & (mag)         & 
(mag)       & (mag) & class      \\
\tableline
\tableline
  V1 &  0$^h$45$^m$59.76$^s$ & 40$^{\rm o}$41$\arcmin$18.2$\arcsec$ & 0.382 & 25.38 & 25.99 & 0.42 & 0.30 & c\\
  V5 &  0$^h$46$^m$ 0.61$^s$ & 40$^{\rm o}$40$\arcmin$52.4$\arcsec$ & 0.339 & 25.55 & 26.17 & 0.40 & 0.29 & c\\
  V8 &  0$^h$46$^m$ 1.02$^s$ & 40$^{\rm o}$40$\arcmin$44.1$\arcsec$ & 0.306 & 25.61 & 26.27 & 0.37 & 0.29 & c\\
 V10 &  0$^h$46$^m$ 0.68$^s$ & 40$^{\rm o}$41$\arcmin$ 2.9$\arcsec$ & 0.687 & 25.17 & 25.72 & 0.97 & 0.66 & ab\\
 V11 &  0$^h$45$^m$56.88$^s$ & 40$^{\rm o}$43$\arcmin$44.1$\arcsec$ & 0.331 & 25.64 & 26.25 & 0.40 & 0.28 & c\\
 V27 &  0$^h$45$^m$57.64$^s$ & 40$^{\rm o}$43$\arcmin$33.5$\arcsec$ & 0.287 & 25.65 & 26.32 & 0.49 & 0.32 & c\\
 V28 &  0$^h$46$^m$ 0.08$^s$ & 40$^{\rm o}$42$\arcmin$ 2.9$\arcsec$ & 0.532 & 25.59 & 26.15 & 0.85 & 0.70 & ab\\
 V36 &  0$^h$46$^m$ 2.04$^s$ & 40$^{\rm o}$41$\arcmin$29.0$\arcsec$ & 0.627 & 25.32 & 25.91 & 1.06 & 0.73 & ab\\
 V37 &  0$^h$46$^m$ 3.26$^s$ & 40$^{\rm o}$40$\arcmin$39.9$\arcsec$ & 0.301 & 25.48 & 26.17 & 0.46 & 0.33 & c\\
 V40 &  0$^h$46$^m$ 1.20$^s$ & 40$^{\rm o}$42$\arcmin$19.8$\arcsec$ & 0.289 & 25.40 & 26.14 & 0.15 & 0.10 & c\\
 V42 &  0$^h$46$^m$ 0.77$^s$ & 40$^{\rm o}$43$\arcmin$ 0.9$\arcsec$ & 0.552 & 25.35 & 25.91 & 1.05 & 0.71 & ab\\
 V43 &  0$^h$46$^m$ 1.74$^s$ & 40$^{\rm o}$42$\arcmin$20.6$\arcsec$ & 0.338 & 25.39 & 26.04 & 0.56 & 0.36 & c\\
 V44 &  0$^h$45$^m$59.93$^s$ & 40$^{\rm o}$43$\arcmin$38.9$\arcsec$ & 0.464 & 25.67 & 26.28 & 1.07 & 0.79 & ab\\
 V47 &  0$^h$46$^m$ 1.55$^s$ & 40$^{\rm o}$42$\arcmin$44.8$\arcsec$ & 0.496 & 25.57 & 26.15 & 1.09 & 0.80 & ab\\
 V50 &  0$^h$46$^m$ 2.68$^s$ & 40$^{\rm o}$42$\arcmin$31.0$\arcsec$ & 0.366 & 25.47 & 26.13 & 0.45 & 0.31 & c\\
 V54 &  0$^h$46$^m$ 4.33$^s$ & 40$^{\rm o}$41$\arcmin$35.2$\arcsec$ & 0.366 & 25.42 & 26.05 & 0.38 & 0.26 & c\\
 V57 &  0$^h$46$^m$ 2.60$^s$ & 40$^{\rm o}$43$\arcmin$10.6$\arcsec$ & 0.554 & 25.42 & 25.99 & 1.03 & 0.68 & ab\\
 V59 &  0$^h$46$^m$ 3.80$^s$ & 40$^{\rm o}$42$\arcmin$36.9$\arcsec$ & 0.339 & 25.46 & 26.12 & 0.43 & 0.30 & c\\
 V66 &  0$^h$46$^m$ 4.71$^s$ & 40$^{\rm o}$42$\arcmin$32.6$\arcsec$ & 0.714 & 25.37 & 25.87 & 0.51 & 0.39 & ab\\
 V76 &  0$^h$46$^m$ 5.86$^s$ & 40$^{\rm o}$42$\arcmin$52.9$\arcsec$ & 0.274 & 25.42 & 26.14 & 0.15 & 0.11 & c\\
 V77 &  0$^h$46$^m$ 4.22$^s$ & 40$^{\rm o}$44$\arcmin$ 5.8$\arcsec$ & 0.678 & 25.52 & 26.04 & 0.63 & 0.43 & ab\\
 V78 &  0$^h$46$^m$ 8.34$^s$ & 40$^{\rm o}$41$\arcmin$24.6$\arcsec$ & 0.774 & 25.38 & 25.92 & 0.49 & 0.42 & ab\\
 V79 &  0$^h$46$^m$ 4.60$^s$ & 40$^{\rm o}$44$\arcmin$ 9.8$\arcsec$ & 0.529 & 25.51 & 26.07 & 1.12 & 0.77 & ab\\
 V80 &  0$^h$46$^m$ 5.78$^s$ & 40$^{\rm o}$43$\arcmin$28.8$\arcsec$ & 0.313 & 25.52 & 26.21 & 0.42 & 0.28 & c\\
 V82 &  0$^h$46$^m$ 5.29$^s$ & 40$^{\rm o}$44$\arcmin$ 2.4$\arcsec$ & 0.735 & 25.48 & 25.99 & 0.47 & 0.37 & ab\\
 V83 &  0$^h$46$^m$ 8.84$^s$ & 40$^{\rm o}$41$\arcmin$35.5$\arcsec$ & 0.351 & 25.36 & 26.03 & 0.52 & 0.35 & c\\
 V88 &  0$^h$46$^m$ 9.86$^s$ & 40$^{\rm o}$41$\arcmin$ 3.5$\arcsec$ & 0.506 & 25.56 & 26.15 & 1.07 & 0.80 & ab\\
 V89 &  0$^h$46$^m$ 6.68$^s$ & 40$^{\rm o}$43$\arcmin$21.9$\arcsec$ & 0.267 & 25.53 & 26.26 & 0.41 & 0.28 & c\\
 V90 &  0$^h$46$^m$ 7.71$^s$ & 40$^{\rm o}$42$\arcmin$51.1$\arcsec$ & 0.353 & 25.49 & 26.11 & 0.50 & 0.26 & d\\
 V95 &  0$^h$46$^m$ 9.60$^s$ & 40$^{\rm o}$41$\arcmin$39.8$\arcsec$ & 0.361 & 25.42 & 26.05 & 0.36 & 0.26 & c\\
V100 &  0$^h$46$^m$ 6.94$^s$ & 40$^{\rm o}$43$\arcmin$58.7$\arcsec$ & 0.275 & 25.59 & 26.29 & 0.48 & 0.35 & c\\
V102 &  0$^h$46$^m$ 8.57$^s$ & 40$^{\rm o}$43$\arcmin$10.5$\arcsec$ & 0.301 & 25.52 & 26.21 & 0.46 & 0.31 & c\\
V112 &  0$^h$46$^m$11.88$^s$ & 40$^{\rm o}$41$\arcmin$21.5$\arcsec$ & 0.573 & 25.41 & 26.00 & 0.95 & 0.62 & ab\\
V114 &  0$^h$46$^m$10.11$^s$ & 40$^{\rm o}$42$\arcmin$54.5$\arcsec$ & 0.726 & 25.30 & 25.80 & 0.33 & 0.24 & ab\\
V120 &  0$^h$46$^m$12.78$^s$ & 40$^{\rm o}$41$\arcmin$24.6$\arcsec$ & 0.283 & 25.39 & 26.11 & 0.52 & 0.35 & c\\
V122 &  0$^h$46$^m$13.10$^s$ & 40$^{\rm o}$41$\arcmin$14.2$\arcsec$ & 0.589 & 25.36 & 25.96 & 1.09 & 0.52 & ab\\
V123 &  0$^h$46$^m$12.46$^s$ & 40$^{\rm o}$41$\arcmin$42.2$\arcsec$ & 0.631 & 25.27 & 25.82 & 0.67 & 0.50 & ab\\
V124 &  0$^h$46$^m$ 9.88$^s$ & 40$^{\rm o}$43$\arcmin$34.7$\arcsec$ & 0.627 & 25.35 & 25.89 & 0.75 & 0.52 & ab\\
V126 &  0$^h$46$^m$10.38$^s$ & 40$^{\rm o}$43$\arcmin$32.5$\arcsec$ & 0.534 & 25.48 & 26.05 & 0.96 & 0.68 & ab\\
V130 &  0$^h$46$^m$11.91$^s$ & 40$^{\rm o}$42$\arcmin$45.1$\arcsec$ & 0.263 & 25.66 & 26.37 & 0.48 & 0.33 & c\\
V131 &  0$^h$46$^m$13.72$^s$ & 40$^{\rm o}$41$\arcmin$30.6$\arcsec$ & 0.327 & 25.50 & 26.15 & 0.38 & 0.22 & c\\
V133 &  0$^h$46$^m$13.43$^s$ & 40$^{\rm o}$41$\arcmin$54.5$\arcsec$ & 0.572 & 25.37 & 25.95 & 0.88 & 0.64 & ab\\
V136 &  0$^h$46$^m$10.23$^s$ & 40$^{\rm o}$44$\arcmin$23.2$\arcsec$ & 0.634 & 25.56 & 26.06 & 0.47 & 0.32 & ab\\
V137 &  0$^h$46$^m$11.13$^s$ & 40$^{\rm o}$43$\arcmin$48.8$\arcsec$ & 0.280 & 25.60 & 26.27 & 0.34 & 0.22 & c\\
V140 &  0$^h$46$^m$14.51$^s$ & 40$^{\rm o}$41$\arcmin$37.0$\arcsec$ & 0.559 & 25.38 & 25.96 & 1.12 & 0.84 & ab\\
V142 &  0$^h$46$^m$11.47$^s$ & 40$^{\rm o}$44$\arcmin$ 4.1$\arcsec$ & 0.554 & 25.41 & 25.97 & 0.91 & 0.65 & ab\\
V147 &  0$^h$46$^m$11.66$^s$ & 40$^{\rm o}$44$\arcmin$23.3$\arcsec$ & 0.442 & 25.52 & 26.10 & 1.12 & 0.68 & ab\\
V157 &  0$^h$46$^m$16.21$^s$ & 40$^{\rm o}$41$\arcmin$47.6$\arcsec$ & 0.329 & 25.47 & 26.16 & 0.40 & 0.29 & c\\
V160 &  0$^h$46$^m$14.30$^s$ & 40$^{\rm o}$43$\arcmin$22.7$\arcsec$ & 0.611 & 25.26 & 25.81 & 0.68 & 0.53 & ab\\
V161 &  0$^h$46$^m$13.32$^s$ & 40$^{\rm o}$44$\arcmin$18.7$\arcsec$ & 0.330 & 25.56 & 26.18 & 0.37 & 0.27 & c\\
V162 &  0$^h$46$^m$16.08$^s$ & 40$^{\rm o}$42$\arcmin$25.4$\arcsec$ & 0.533 & 25.57 & 26.15 & 1.01 & 0.78 & ab\\
V163 &  0$^h$46$^m$15.92$^s$ & 40$^{\rm o}$42$\arcmin$37.9$\arcsec$ & 0.313 & 25.35 & 26.00 & 0.37 & 0.25 & c\\
V164 &  0$^h$46$^m$14.21$^s$ & 40$^{\rm o}$43$\arcmin$55.8$\arcsec$ & 0.580 & 25.45 & 26.01 & 0.85 & 0.64 & ab\\
V166 &  0$^h$46$^m$13.71$^s$ & 40$^{\rm o}$44$\arcmin$24.4$\arcsec$ & 0.583 & 25.46 & 25.99 & 0.65 & 0.59 & ab\\
V167 &  0$^h$46$^m$10.38$^s$ & 40$^{\rm o}$43$\arcmin$44.0$\arcsec$ & 0.621 & 25.43 & 25.95 & 0.67 & 0.47 & ab\\
\tableline
\end{tabular}
\end{center}
}

\newpage

\parbox{3.25in}{
\begin{center}
{\sc Table 2\\ RR Lyrae Photometry$\tablenotemark{a}$}

\begin{tabular}{ccccc}
\tableline
MJD & & \multicolumn{3}{c}{Photometry (mag)$\tablenotemark{b}$} \\
(days) & Band & V1 & V5 & V8 \\
\tableline
\tableline
52610.06377 & F606W & 25.58 & 25.75 & 25.87 \\
52610.12537 & F606W & 25.27 & 25.57 & 25.57 \\
52610.19249 & F606W & 99.99 & 25.42 & 25.48 \\
52610.25919 & F606W & 25.22 & 25.44 & 25.51 \\
52611.20557 & F814W & 26.09 & 26.14 & 26.39 \\
52611.25942 & F814W & 25.98 & 26.07 & 99.99 \\
52611.27610 & F814W & 25.96 & 26.09 & 26.47 \\
52611.32653 & F814W & 25.75 & 26.19 & 26.37 \\
52611.39623 & F606W & 25.29 & 25.74 & 25.42 \\
52611.41250 & F606W & 25.37 & 25.83 & 25.47 \\
\tableline
\end{tabular}
\end{center}
{\small $^{\rm a}$Table 2 is published in its entirety in the electronic 
edition of The Astronomical Journal. A portion is shown here for guidance 
regarding its form and content.\\
$^{\rm b}$A magnitude of 99.99 signifies missing data.}
}

\vskip 0.1in

\noindent
with respect to the Galactic halo,
therefore resulting in different age vs. chemical composition patterns.

\subsection{Specific Frequency}

The specific frequency of RR Lyraes in the M31 halo has been the
subject of debate, due perhaps in part 
to the difficulty of observing these stars from
the ground.  Pritchet \& van den Bergh (1987) observed a
$2.2\arcmin \times 3.3\arcmin$ field 40$\arcmin$ from the nucleus on
the southeast minor axis, and found 30 RR Lyraes with amplitudes
$\gtrsim$0.7~mag.  Assuming a completeness of 25\%, they determined
there should be 120 RR Lyraes with amplitudes $\gtrsim$0.7~mag.  They
took this to be a very strict lower limit, given the lack of
low-amplitude RR Lyraes, and claimed their value was 55\% of the
frequency for M3 (NGC5272), a Galactic  globular cluster with a high
frequency of RR Lyraes.  The 
\linebreak 

\hspace{0.0in}
\parbox{6.5in}{
\begin{center}
{\sc Table 3\\ Positions of Other Variables}

\begin{tabular}{ccccccccc}
\tableline
     &     R.A.     & Dec.         &      &     R.A.     & Dec.         &      &     R.A.     & Dec.         \\
Name &      (J2000) &      (J2000) & Name &      (J2000) &      (J2000) & Name &      (J2000) &      (J2000) \\
\tableline
\tableline
  V6 &  0$^h$45$^m$57.91$^s$ & 40$^{\rm o}$42$\arcmin$48.1$\arcsec$ &  V64 &  0$^h$46$^m$ 3.27$^s$ & 40$^{\rm o}$43$\arcmin$29.4$\arcsec$ & V118 &  0$^h$46$^m$10.05$^s$ & 40$^{\rm o}$43$\arcmin$ 6.0$\arcsec$ \\
  V7 &  0$^h$45$^m$58.14$^s$ & 40$^{\rm o}$42$\arcmin$38.6$\arcsec$ &  V65 &  0$^h$46$^m$ 4.06$^s$ & 40$^{\rm o}$42$\arcmin$59.2$\arcsec$ & V119 &  0$^h$46$^m$11.23$^s$ & 40$^{\rm o}$42$\arcmin$19.0$\arcsec$ \\
 V12 &  0$^h$45$^m$58.06$^s$ & 40$^{\rm o}$42$\arcmin$55.2$\arcsec$ &  V67 &  0$^h$46$^m$ 4.42$^s$ & 40$^{\rm o}$42$\arcmin$46.9$\arcsec$ & V127 &  0$^h$46$^m$11.02$^s$ & 40$^{\rm o}$43$\arcmin$ 5.9$\arcsec$ \\
 V14 &  0$^h$45$^m$58.78$^s$ & 40$^{\rm o}$42$\arcmin$29.7$\arcsec$ &  V70 &  0$^h$46$^m$ 3.88$^s$ & 40$^{\rm o}$43$\arcmin$28.5$\arcsec$ & V128 &  0$^h$46$^m$12.17$^s$ & 40$^{\rm o}$42$\arcmin$18.4$\arcsec$ \\
 V18 &  0$^h$45$^m$58.87$^s$ & 40$^{\rm o}$42$\arcmin$32.3$\arcsec$ &  V72 &  0$^h$46$^m$ 6.25$^s$ & 40$^{\rm o}$41$\arcmin$53.8$\arcsec$ & V129 &  0$^h$46$^m$ 9.88$^s$ & 40$^{\rm o}$43$\arcmin$57.4$\arcsec$ \\
 V19 &  0$^h$45$^m$58.90$^s$ & 40$^{\rm o}$42$\arcmin$30.9$\arcsec$ &  V74 &  0$^h$46$^m$ 4.44$^s$ & 40$^{\rm o}$43$\arcmin$25.7$\arcsec$ & V134 &  0$^h$46$^m$13.50$^s$ & 40$^{\rm o}$41$\arcmin$53.2$\arcsec$ \\
 V20 &  0$^h$46$^m$ 1.02$^s$ & 40$^{\rm o}$41$\arcmin$ 2.1$\arcsec$ &  V84 &  0$^h$46$^m$ 5.30$^s$ & 40$^{\rm o}$44$\arcmin$ 5.8$\arcsec$ & V139 &  0$^h$46$^m$11.90$^s$ & 40$^{\rm o}$43$\arcmin$17.9$\arcsec$ \\
 V21 &  0$^h$45$^m$58.90$^s$ & 40$^{\rm o}$42$\arcmin$32.7$\arcsec$ &  V85 &  0$^h$46$^m$ 9.12$^s$ & 40$^{\rm o}$41$\arcmin$24.7$\arcsec$ & V143 &  0$^h$46$^m$14.94$^s$ & 40$^{\rm o}$41$\arcmin$42.8$\arcsec$ \\
 V25 &  0$^h$45$^m$58.97$^s$ & 40$^{\rm o}$42$\arcmin$32.0$\arcsec$ &  V86 &  0$^h$46$^m$ 9.55$^s$ & 40$^{\rm o}$41$\arcmin$ 8.5$\arcsec$ & V145 &  0$^h$46$^m$14.38$^s$ & 40$^{\rm o}$42$\arcmin$13.8$\arcsec$ \\
 V29 &  0$^h$46$^m$ 1.76$^s$ & 40$^{\rm o}$40$\arcmin$53.1$\arcsec$ &  V91 &  0$^h$46$^m$ 7.74$^s$ & 40$^{\rm o}$42$\arcmin$50.6$\arcsec$ & V150 &  0$^h$46$^m$12.63$^s$ & 40$^{\rm o}$43$\arcmin$48.2$\arcsec$ \\
 V31 &  0$^h$46$^m$ 2.46$^s$ & 40$^{\rm o}$40$\arcmin$41.3$\arcsec$ &  V93 &  0$^h$46$^m$ 9.40$^s$ & 40$^{\rm o}$41$\arcmin$45.2$\arcsec$ & V151 &  0$^h$46$^m$13.16$^s$ & 40$^{\rm o}$43$\arcmin$36.0$\arcsec$ \\
 V38 &  0$^h$46$^m$ 1.79$^s$ & 40$^{\rm o}$41$\arcmin$44.0$\arcsec$ &  V94 &  0$^h$46$^m$ 9.77$^s$ & 40$^{\rm o}$41$\arcmin$29.9$\arcsec$ & V158 &  0$^h$46$^m$13.67$^s$ & 40$^{\rm o}$43$\arcmin$37.6$\arcsec$ \\
 V39 &  0$^h$46$^m$ 0.08$^s$ & 40$^{\rm o}$42$\arcmin$57.0$\arcsec$ &  V99 &  0$^h$46$^m$ 6.54$^s$ & 40$^{\rm o}$44$\arcmin$ 8.0$\arcsec$ & V159 &  0$^h$46$^m$14.91$^s$ & 40$^{\rm o}$42$\arcmin$51.6$\arcsec$ \\
 V41 &  0$^h$46$^m$ 1.07$^s$ & 40$^{\rm o}$42$\arcmin$44.9$\arcsec$ & V103 &  0$^h$46$^m$ 8.70$^s$ & 40$^{\rm o}$43$\arcmin$10.2$\arcsec$ & V168 &  0$^h$46$^m$ 0.63$^s$ & 40$^{\rm o}$41$\arcmin$13.8$\arcsec$ \\
 V48 &  0$^h$46$^m$ 2.27$^s$ & 40$^{\rm o}$42$\arcmin$16.3$\arcsec$ & V105 &  0$^h$46$^m$ 9.32$^s$ & 40$^{\rm o}$42$\arcmin$53.2$\arcsec$ & V169 &  0$^h$46$^m$ 2.52$^s$ & 40$^{\rm o}$40$\arcmin$37.7$\arcsec$ \\
 V51 &  0$^h$46$^m$ 1.59$^s$ & 40$^{\rm o}$43$\arcmin$18.1$\arcsec$ & V106 &  0$^h$46$^m$10.84$^s$ & 40$^{\rm o}$41$\arcmin$51.5$\arcsec$ & V170 &  0$^h$46$^m$ 8.72$^s$ & 40$^{\rm o}$41$\arcmin$53.0$\arcsec$ \\
 V52 &  0$^h$46$^m$ 2.51$^s$ & 40$^{\rm o}$42$\arcmin$39.2$\arcsec$ & V111 &  0$^h$46$^m$11.28$^s$ & 40$^{\rm o}$41$\arcmin$42.7$\arcsec$ & V171 &  0$^h$46$^m$ 7.35$^s$ & 40$^{\rm o}$44$\arcmin$17.1$\arcsec$ \\
 V60 &  0$^h$46$^m$ 4.91$^s$ & 40$^{\rm o}$41$\arcmin$55.9$\arcsec$ & V113 &  0$^h$46$^m$ 8.95$^s$ & 40$^{\rm o}$43$\arcmin$29.5$\arcsec$ & V172 &  0$^h$46$^m$ 9.93$^s$ & 40$^{\rm o}$43$\arcmin$18.3$\arcsec$ \\
 V61 &  0$^h$46$^m$ 5.72$^s$ & 40$^{\rm o}$41$\arcmin$24.2$\arcsec$ & V116 &  0$^h$46$^m$12.44$^s$ & 40$^{\rm o}$41$\arcmin$20.8$\arcsec$ & V173 &  0$^h$46$^m$10.45$^s$ & 40$^{\rm o}$43$\arcmin$ 1.6$\arcsec$ \\
 V63 &  0$^h$46$^m$ 5.43$^s$ & 40$^{\rm o}$41$\arcmin$55.9$\arcsec$ & V117 &  0$^h$46$^m$11.67$^s$ & 40$^{\rm o}$41$\arcmin$54.6$\arcsec$ & V174 &  0$^h$46$^m$ 8.98$^s$ & 40$^{\rm o}$43$\arcmin$ 9.2$\arcsec$ \\
\tableline
\end{tabular}
\end{center}
}

\noindent
specific frequency of RR Lyraes in M3, as
given by Harris (1996) and normalized to a total cluster luminosity
($M_{Vt}$) of $-7.5$~mag, is $S_{RR} = N_{RR} 10^{(M_{Vt}+7.5)/2.5} =
49$.  Thus, Pritchet \& van den Bergh (1987) determined $S_{RR} \approx
27$, which was a surprisingly high number, given the mean metallicity
[Fe/H]$=-0.6$ of Mould \& Kristian (1986).

Recently, Dolphin et al.\ (2003) observed a $9.6\arcmin \times
9.6\arcmin$ field that included the Pritchet \& van den Bergh (1987)
field, and found a much lower frequency of RR Lyraes.  After finding
24 RR Lyraes and estimating a completeness of 24\%, they estimated
that there are 100 RR Lyraes in their field.  Given the difference in
field size, the Dolphin et al.\ (2003) frequency is 15 times smaller
than that of Pritchet \& van den Bergh (1987), i.e., $S_{RR} \approx
1.8$.  Dolphin et al.\ (2003) claimed that their lower frequency was
supported by the CMD of Brown et al.\ (2003) -- the same data we are
using in this paper -- because they estimated that there were only 10
RR Lyraes in the Brown et al.\ (2003) data.  However, the CMD shown by
Brown et al.\ (2003) was displayed as a greyscale Hess diagram, in
order to clearly show the characteristics of the main sequence turnoff
in a catalog of $\sim 300,000$ stars; individual stars were not shown,
nor was variability indicated in any way.  In reality, 41 of the 55 RR
Lyraes reported in our current work also appeared in the CMD of Brown
et al.\ (2003); the remaining 14 were in the $\sim 20$\% of the image area
masked for that earlier analysis of the star formation history (see $\S$2).

These earlier estimates of the RR Lyrae frequency were limited by
several factors.  The first was uncertainty in the completeness.  RR
Lyraes were at the edge of detection in both the Dolphin et al.\
(2003) and Pritchet \& van den Bergh (1987) studies, and neither group
quantified the completeness at that depth through artificial star
tests.  Furthermore, the total luminosity in these fields could 
be estimated only from photographic plates or counts of the bright RGB
stars.  In contrast, our artificial star tests indicate that the
completeness in our data \linebreak

\newpage

\hskip 0.2in
\epsfxsize=6.5in \epsfbox{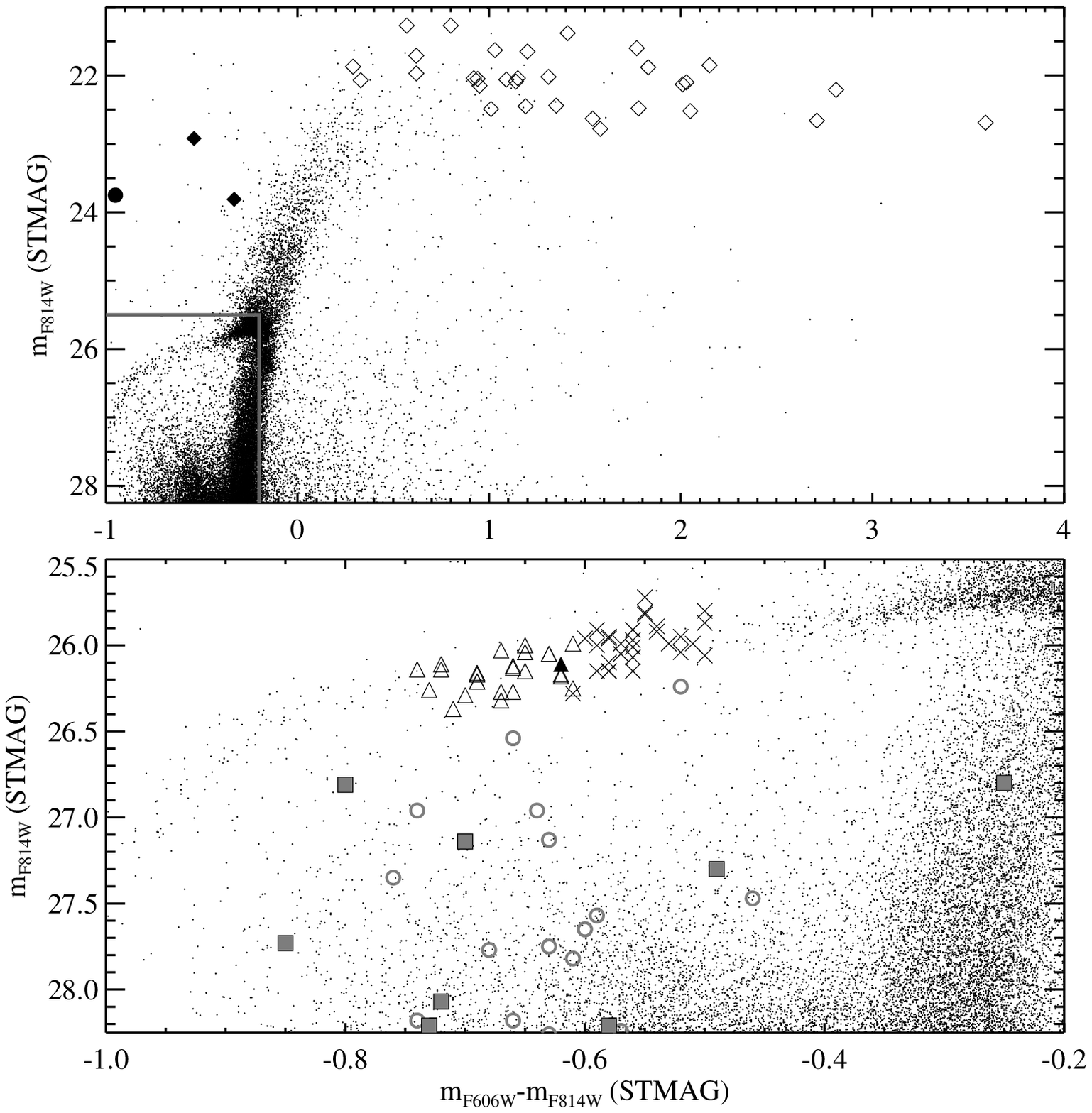}

\hskip 0.2in
\parbox{6.5in}{\small {\sc Fig.~5--}
{\it Top panel:}
Subset of the M31 halo CMD of Brown et al.\ (2003), highlighting the
bright variables.  From Figure 2, V84 is marked by a closed circle, while
V118 and V60 are marked by closed diamonds.  The variables of Figure 4
are shown by open diamonds.  {\it Bottom panel:}  Expanded view of the
faint variables: RRab stars ({\it crosses}), RRc stars ({\it open triangles}),
an RRd star ({\it closed triangle}), dwarf Cepheids ({\it open circles}),
and eclipsing binaries ({\it closed squares}).}

\vskip 0.12in

\noindent
is $\approx$100\% at the horizontal branch,
our time sampling enables a search for periodicity with high
statistical significance, and our deep catalog allows an accurate
determination of the total luminosity in our field.  Brown et al.\
(2003) determined $\mu_V \approx 26.3$~mag arcsec$^{-2}$ in this field.
Assuming an extinction of $E(B-V) = 0.08 \pm 0.03$~mag (Schlegel et
al.\ 1998) and a distance modulus $(m-M)_0 = 24.44\pm 0.1$~mag
(Freedman \& Madore 1990), we estimate a total brightness in our field
of $M_V \approx -10.0$~mag, and $S_{RR} \approx 5.6$ 
($\sim$ 1 RR Lyrae star per $1.5 \times 10^4 L_\odot$).  Our frequency
is $\sim$3 times higher than that found by Dolphin et al.\ (2003), with
the number of RR Lyraes compared either to the number of RGB stars 
(the metric of Dolphin et al.\ 2003) or the
total luminosity of the field (the metric of Harris 1996).
The specific frequency we find is much
higher than one would expect for a population with metallicity
[Fe/H]$=-0.8$ (Figure 10), but not surprising given that the \linebreak

\vskip 6.97in

\noindent
M31 halo
has a wide metallicity distribution, with $\sim$40\% of the stars at
$-2.5<$[Fe/H]$ < -1$ (Durrell et al.\ 2001).  
Because about half of the M31 halo is
metal-rich and of intermediate age (Brown et al.\ 2003), and thus
incapable of producing RR Lyrae stars, one could
reasonably shift the location of the M31 data point in Figure 10 to a
metallicity near [Fe/H]$=-1.6$ and to a frequency twice as large
as that found at ${\rm [Fe/H]}=-0.8$.

\subsection{Metallicity}

Given a larger number of RR Lyraes than expected for a mean
metallicity of [Fe/H]$=-0.8$, it is reasonable to assume that the RR
Lyraes come from the old, metal-poor component of the M31 halo.  As
done in studies of dSph populations (e.g., Pritzl et al.\ 2002; Siegel
\& Majewski 2000), we can use one of several empirical relations to
estimate the metallicity of the RR Lyraes in our field.  Looking at
the mean properties of the \linebreak

\vskip -0.18in
\epsfxsize=3.15in \epsfbox{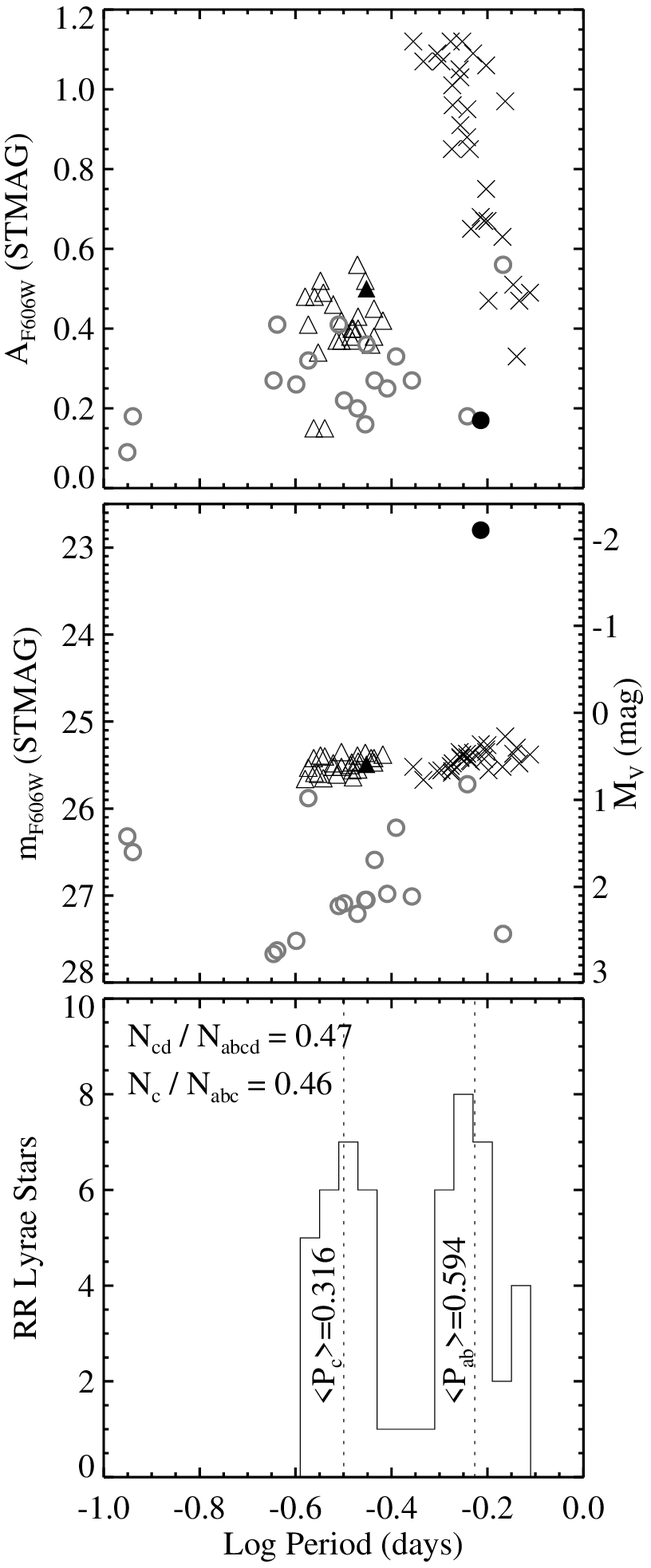}

\noindent
\parbox{3.4in}{\small {\sc Fig.~6--}
{\it Top panel:} The amplitude vs. period diagram
for short period variables (excluding the eclipsing binaries).
The RRab stars ({\it crosses}) are well-separated from the
RRc stars ({\it open triangles}), but the dwarf Cepheids ({\it open
circles}) are mixed in with the RR Lyraes.
The RRd star ({\it closed triangle})
is mixed in the main RRc clump.  The bright blue
variable above the HB is also marked ({\it closed circle}).
{\it Middle panel:} The luminosity vs. period diagram, with
the same symbols.  The RR Lyraes are well-separated from the
other variables, although two ``dwarf Cepheids'' might actually
be RR Lyraes.  The left axis provides the observed F606W magnitudes, while
the right gives the transformation to $M_V$ appropriate for the color
of the center of the RR Lyrae gap, assuming the Cepheid distance to M31
(see $\S$4.6 for details).
{\it Bottom panel:} The period histogram
for the RR Lyraes.  The parameters used to distinguish
Oosterhoff type are labeled.}

\epsfxsize=3.25in \epsfbox{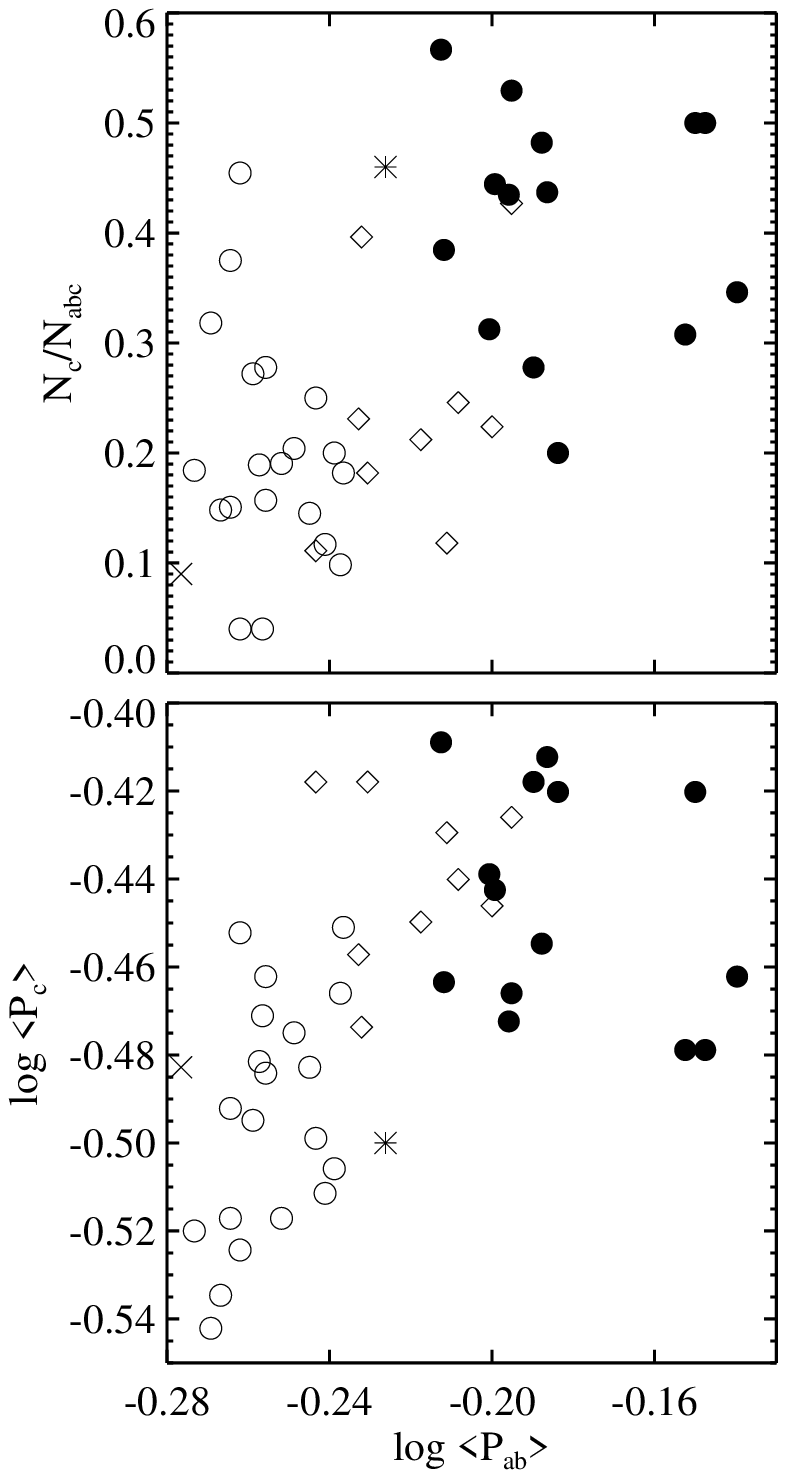}

\noindent
\parbox{3.4in}{\small {\sc Fig.~7--}
{\it Top panel:} The fractions of RRab+RRc stars that
are RRc, vs. the mean RRab periods, for different populations.
Galactic globular clusters (Clement et al.\ 2001)
of Oosterhoff I ({\it open circles}) and
Oosterhoff II ({\it closed circles}) types are well separated, while
Local Group dSphs ({\it diamonds}; Pritzl et al.\ 2002, 2004;
Dall'Ora et al.\ 2003) and the M31 halo ({\it asterisk})
bridge the gap; the Milky Way halo field is, on average, Oosterhoff I
({\it cross}; Cacciari \& Renzini 1976).
The fraction of RRc stars in the M31
halo is like that in Oosterhoff II clusters, while the fraction of RRc
stars in dSphs is like that in Oosterhoff I clusters.  {\it Bottom
panel:} The mean RRc period vs. the mean RRab period.  Again, the
dSphs and the M31 halo bridge the gap between Oosterhoff types, but
the mean RRc period in M31 is like that in Oosterhoff I clusters,
while the mean RRc period in dSphs is like that in Oosterhoff II
clusters.}

\newpage

\epsfxsize=3.25in \epsfbox{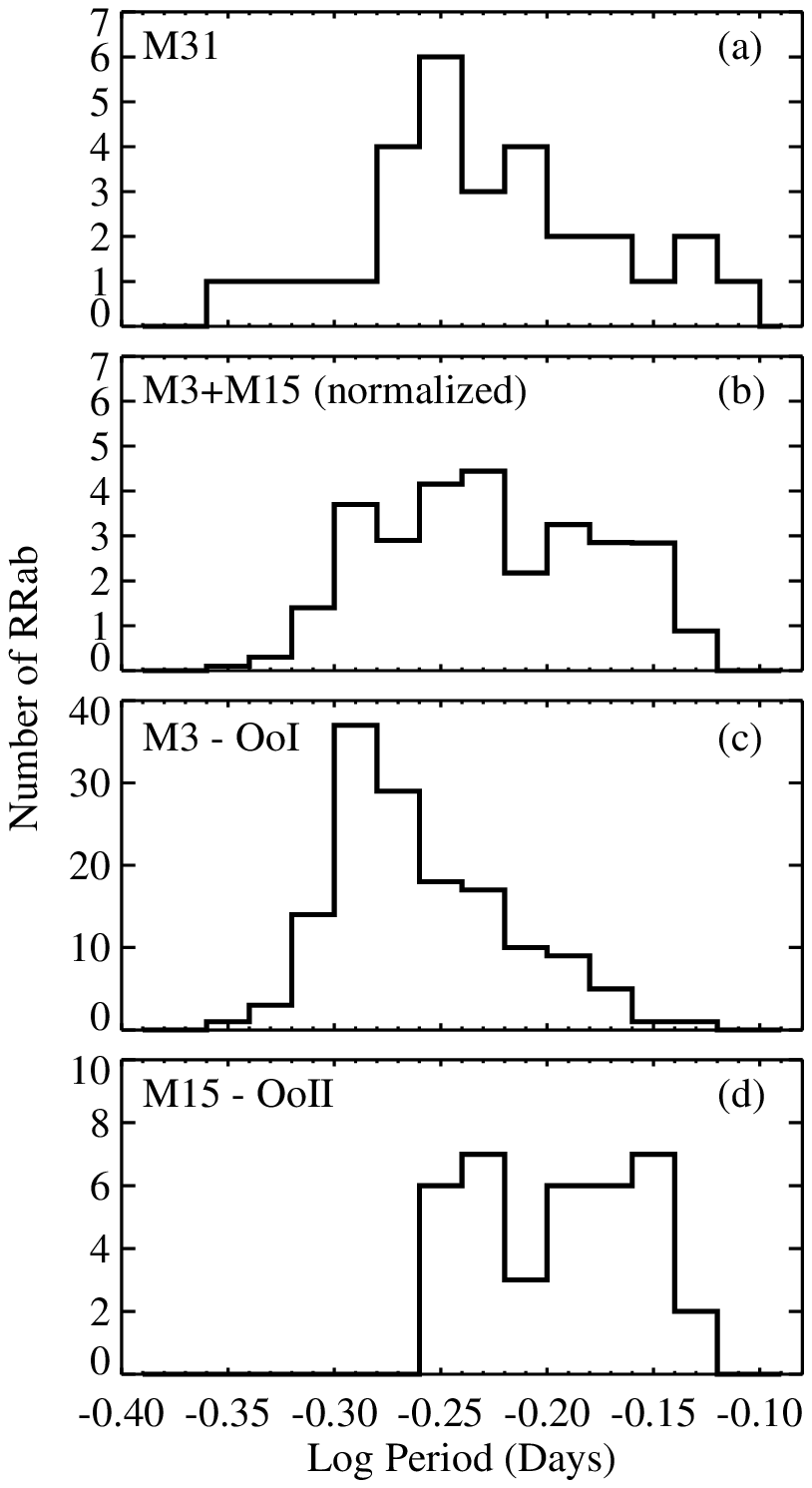}

\noindent
\parbox{3.4in}{\small {\sc Fig.~8--}
{\it Panel (a):} The period distribution
of RRab stars in the M31 halo.  {\it Panel (b):} The period
distribution of RRab stars when the stars are drawn equally
from M3 and M15 -- two clusters rich in RR Lyraes with distinct
Oosterhoff types.  {\it Panel (c):} The period distribution in M3
(Clement et al.\ 2001).
{\it Panel (d):} The period distribution in M15 (Clement et al.\ 2001).}

\vskip 0.2in

\noindent
RRab and RRc stars, the period-metallicity
relations of Sandage (1993) give
[Fe/H]~=~(--log$<$$P_{ab}$$>-0.389)/0.092 = -1.77$, and
[Fe/H]~=~(--log$<$$P_c$$> -0.670)/0.119 = -1.43$, on the Zinn \& West
(1984) scale.  Given the ensemble population in the M31 halo and the
scatter for these relations shown by Sandage (1993), the difference in
metallicity for the RRab and RRc stars is not significant, but we can
say that the RR Lyraes lie in the metal-poor tail of the halo
metallicity distribution.  This finding is further supported by the
period-amplitude-metallicity relation of Alcock et al.\ (2000), which
provides metallicities on the Zinn \& West (1984) scale for individual
RRab stars: [Fe/H]=~--8.85(log$P_{ab}+0.15 A_V$)--2.60.  Converting
our amplitudes in the $m_{F606W}$ bandpass to those in Johnson $V$, we
find a mean [Fe/H] of --1.79, with a standard deviation of 0.32; the
accuracy of this method is $\sigma_{\rm [Fe/H]}=0.31$ per star (Alcock 
et al.\ 2000), so the dispersion in our individual metallicities 
is not statistically significant.

\epsfxsize=3.25in \epsfbox{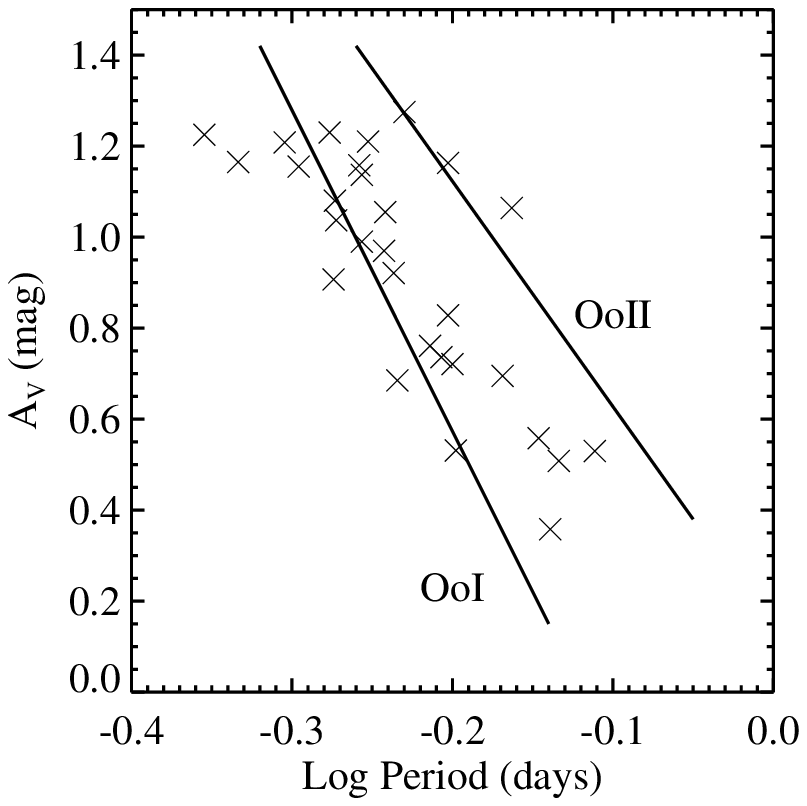}

\noindent
\parbox{3.5in}{\small {\sc Fig.~9--}
The period-amplitude diagram for the RRab stars
of the M31 halo ({\it crosses}), compared to the period-amplitude
relations for Oosterhoff I and Oosterhoff II clusters ({\it lines};
Clement 2000).  Note that we converted
amplitudes in the $m_{F606W}$ bandpass
to amplitudes in Johnson $V$ using Lejeune synthetic spectra, but
this correction is small (a 6--10\% increase in the F606W amplitude).
The M31 halo stars appear to be an intermediate
case between Oosterhoff types, although there is a tendency toward
the Oosterhoff I type at large amplitudes.}

\subsection{Double-Mode Pulsators}

As mentioned in $\S$2, five RR Lyrae stars in our sample show
significant scatter in their light curves, as expected for double-mode
pulsators (RRd stars) and RRab stars exhibiting the Blazhko effect.  
One of these stars, V90, shows clear periodicity at
two different periods, with $P_0=0.4735$ days and $P_1 = 0.3534$ days.
The ratio $P_1/P_0$ is sensitive to the mass of the RRd star,
with RRd stars from the two Oosterhoff types clearly separated in a
$P_1/P_0$ vs $P_0$ diagram (see Bono et al.\ 1996 and references
therein).  Using these periods to place our confirmed RRd star in the
diagrams of Bono et al.\ (1996), we find that the $P_0$ for this star
is similar to that of RRd stars found in Oosterhoff I clusters, but
the $P_1/P_0$ ratio is typical for RRd stars found in Oosterhoff II
clusters, with a mass of $\approx 0.75$~$M_\odot$.  It is curious that
an individual star in the M31 halo cannot be placed into either
Oosterhoff type in such a diagram; again, this suggests that the M31
halo population is not a simple mix of Oosterhoff types.

\subsection{Distance}

With the metallicities derived above, we can estimate the distance to
the M31 halo population using the relation of Carretta et al.\ (2000):
$M_V = (0.18 \pm 0.09)({\rm [Fe/H]}+1.5)+(0.57\pm0.07)$.  Note that
this relation is consistent with the average of the methods
discussed by Cacciari \& Clementini (2003), who give an excellent
review of distance determination using RR Lyrae stars.  Using the
synthetic spectra of Lejeune et al.\ (1997), an assumed extinction of
$E(B-V)=0.08\pm0.03$ (Schlegel et al.\ 1998), and the extinction curve
of Fitzpatrick (1999), we can calculate the offset between $m_{F606W}$
and $V$ as a function of $m_{F606W}-m_{F814W}$.  The RRab stars have a
color range $-0.61 \le (m_{F606W}-m_{F814W}) \le -0.5$~mag, with
corresponding corrections of $-0.17 \le (V-m_{F606W}) \le -0.13$~mag;
the \linebreak

\epsfxsize=3.25in \epsfbox{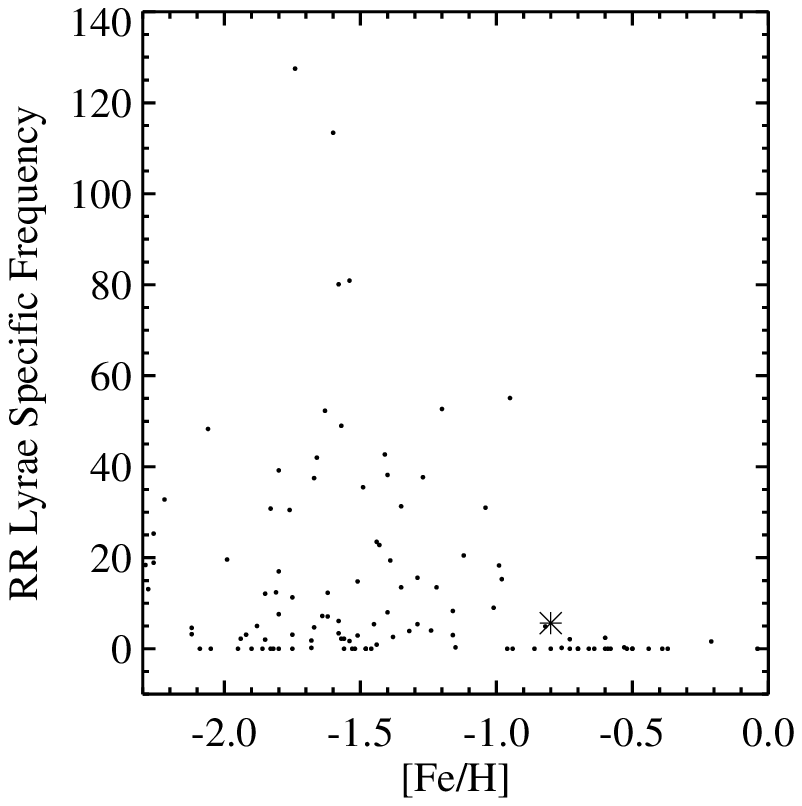}

\noindent
\parbox{3.4in}{\small {\sc Fig.~10--}
The specific frequency of RR Lyraes, normalized to
a population at $M_{Vt} = -7.5$~mag, as a function of metallicity,
for Galactic globular clusters ({\it points}; Harris 1996)
and the M31 halo ({\it asterisk}).
Although the frequency of RR Lyraes in the M31 halo
is higher than expected for the mean halo metallicity, the RR Lyraes
have a metallicity near [Fe/H]$=-1.6$.  Considered as part of the
old metal-poor population, the frequency in the M31 halo is not surprising.
}

\vskip 0.4in

\noindent
RRc stars have a color range $-0.74 \le (m_{F606W}-m_{F814W}) \le
-0.61$~mag, with corresponding corrections of $-0.21 \le (V-m_{F606W})
\le -0.17$~mag.  After correcting the individual stars with the
offsets appropriate to their colors, the RRab stars lie at
$<$$m_{F606W}$$> = 25.43$~mag and $<$$V$$>=25.28\pm 0.01$~mag, while
the RRc stars lie at $<$$m_{F606W}$$> = 25.49$~mag and $<$$V$$>=25.31
\pm 0.01$~mag (the statistical errors on $<$$V$$>$ are much smaller
than 0.01~mag, but the throughputs of the ACS filters are only known
to $\sim$1\%).  Assuming [Fe/H]$=-1.77$, the Carretta et al.\ (2000)
relation yields $M_V = 0.52\pm 0.07$~mag for the RRab stars; assuming
[Fe/H]$=-1.43$ gives $M_V = 0.58\pm 0.07$~mag for the RRc stars.  With
an extinction of $A_V = 0.25 \pm 0.09$, we have distance moduli of
$(m-M)_0 = 24.51\pm 0.11$~mag for the RRab stars, and $(m-M)_0 =24.48
\pm 0.11$~mag for the RRc stars.  Averaging these results gives an RR
Lyrae distance modulus of $(m-M)_0 = 24.5\pm 0.1$~mag, which is in
very good agreement with the Cepheid distance modulus of $(m-M)_0 =
24.44\pm 0.1$~mag (Freedman \& Madore 1990), given the uncertainties.

Dolphin et al.\ (2003) made similar estimates of the RR Lyrae
luminosity in their field, but found $<$$V_0$$> = 24.81\pm 0.11$~mag --
about 0.24~mag brighter than our own estimate.  Given their relatively
low completeness, it is plausible that their RR Lyrae sample is biased
toward brighter stars (although they argue otherwise).  Dolphin
inspected the Brown et al.\ (2003) CMD and found support for their
bright RR Lyrae estimate, but we stress again that our previous
presentation of the M31 halo data only showed a greyscale Hess diagram, making
such estimates from the published figure difficult.

\section{Other Variables}

Although the focus of this paper is the RR Lyrae population in the
Andromeda halo, we briefly note the other classes of variables
suitable for followup studies.  The short period variables below the
HB (see Figures 2 and 5) fall in the broad category of dwarf Cepheids;
given the wide age range in the M31 halo (Brown et al.\ 2003), these
are probably a mix of pulsating and eclipsing stars on the main
sequence (e.g., $\delta$ Scuti stars) and in the blue straggler
population (e.g., SX Phoenicis stars).  Above the HB, the bluest star
(V84) is outside of the Cepheid instability strip, but might be a
pulsating post-asymptotic branch star.  The other two stars above the
HB (V118 and V60) have periods that are longer than typically found
for Anomalous Cepheids, but their luminosities and periods put them
close to the Population I and Population II Cepheids.  In any case, if
they are both pulsating variables, they cannot belong to the same
class, given that the star with the longer period is fainter.  
The semiregulars
and long period variables (see Figure 4) include RV Tauri stars and Mira
stars.  In globular clusters, Miras are found only at [Fe/H]~$>-1$
(Frogel \& Whitelock 1998), where they have periods $\lesssim 310$
days.  Given the wide age spread in the M31 halo, one might expect
some of the Miras to have periods longer than 310 days, but
characterization of these stars would require a longer baseline than
that currently available.

\section{Summary}

We have presented a complete survey of the RR Lyrae stars in an M31
halo field, 51 arcmin from the nucleus.  We find 29 RRab stars with a
mean period of 0.594 days, 25 RRc stars with a mean period of 0.316
days, and 1 RRd star with a fundamental period of 0.473 days and a
first overtone period of 0.353 days.  The RR Lyrae population of the
M31 halo cannot be clearly placed into either Oosterhoff type, and is
distinct from the Milky Way cluster and halo field populations.  In a
broad sense, the Local Group dSphs share the intermediate Oosterhoff
status found the M31 halo, but the characteristics of the M31 RR
Lyraes ($<$$P_{ab}$$>$, $<$$P_{c}$$>$, $N_c / N_{abc}$) are distinct
from the dSphs, suggesting that the M31 halo is not comprised of
dissolved globular clusters like those in the Milky Way or dissolved
Local Group dSphs.  The specific frequency of RR Lyraes ($S_{RR} =
5.6$) is very high for a mean halo metallicity of [Fe/H]$=-0.8$, but
within the normal range when considered as a component of the old,
metal-poor halo population.  The mean metallicity of the RR Lyrae
population is indeed much lower than that of the halo, with a mean
[Fe/H]$=-1.77$ for the RRab stars and a mean [Fe/H]$=-1.43$ for the
RRc stars.  The distance to M31 determined from the RR Lyrae
luminosity is $(m-M)_o = 24.5 \pm 0.1$~mag, in good agreement with the
Cepheid distance.

\acknowledgements

Support for proposal 9453 was provided by NASA through a grant from
STScI, which is operated by AURA, Inc., under NASA contract NAS
5-26555.  We are grateful to P.\ Stetson for providing the latest
version of DAOPHOT and assistance with its use, and to R. Gilliland
for useful discussions.  H. Smith kindly reviewed a draft of this
manuscript and offered helpful comments.  We thank the members of the
scheduling and operations teams at STScI (especially P.\ Royle, D.\
Taylor, and D.\ Soderblom) for their efforts in executing a large
program during a busy HST cycle.  Our paper was improved by the
suggestions of an anonymous referee.

\end{document}